\documentclass{emulateapj}
\usepackage{apjfonts}

\newcommand{\kh}{KH~15D}
\def\ltsima{$\; \buildrel < \over \sim \;$}
\def\lsim{\lower.5ex\hbox{\ltsima}}
\def\gtsima{$\; \buildrel > \over \sim \;$}
\def\gsim{\lower.5ex\hbox{\gtsima}}

\slugcomment{To appear in the Astrophysical Journal}
\shorttitle{Models of \kh}
\shortauthors{Winn et al.}

\begin{document}

\title{ The Orbit and Occultations of KH~15D }

\author{
Joshua N.\ Winn\altaffilmark{1,2,3},
Catrina M.\ Hamilton\altaffilmark{4,5},
William J.\ Herbst\altaffilmark{6},\\
Jennifer L.\ Hoffman\altaffilmark{7},
Matthew J.\ Holman\altaffilmark{1},
John A.\ Johnson\altaffilmark{7},
Marc J.\ Kuchner\altaffilmark{8}
}

\altaffiltext{1}{Harvard-Smithsonian Center for Astrophysics, 60
Garden Street, Mail Stop 51, Cambridge, MA 02138}

\altaffiltext{2}{Hubble Fellow}

\altaffiltext{3}{{\it Current address}: Department of Physics,
  Massachusetts Institute of Technology, 77 Massachusetts Ave.,
  Cambridge, MA 02139}

\altaffiltext{4}{Mount Holyoke College, 50 College Street, South
Hadley, MA 01075}

\altaffiltext{5}{Five College Astronomy Department, University of
Massachusetts, Amherst, MA 01003}

\altaffiltext{6}{Van Vleck Observatory, Wesleyan University,
Middletown, CT 06459}

\altaffiltext{7}{Department of Astronomy, University of California,
Mail Code 3411, Berkeley, CA 94720}

\altaffiltext{8}{NASA Goddard Space Flight Center, Greenbelt, MD
20771}

\begin{abstract}
The unusual flux variations of the pre--main-sequence binary star
KH~15D have been attributed to occultations by a circumbinary disk.
We test whether or not this theory is compatible with newly available
data, including recent radial velocity measurements, CCD photometry
over the past decade, and photographic photometry over the past 50
years. We find the model to be successful, after two refinements: a
more realistic motion of the occulting feature, and a halo around each
star that probably represents scattering by the disk. The occulting
feature is exceptionally sharp-edged, raising the possibility that the
dust in the disk has settled into a thin layer, and providing a tool
for fine-scale mapping of the immediate environment of a T Tauri
star. However, the window of opportunity is closing, as the currently
visible star may be hidden at all orbital phases by as early as 2008.
\end{abstract}


\section{Introduction}
\label{sec:introduction}

After a newborn low-mass star emerges from the dust of its parent
molecular cloud, it spends a few million years as a T~Tauri star. This
phase is characterized by optical variability and chromospheric
emission lines, as the star accretes gas, propels a bipolar outflow,
and contracts onto the main sequence (see, e.g., the review by Bertout
1989). A conventional boundary line is drawn between ``classical'' and
``weak-lined'' T Tauri stars, depending on whether the equivalent
width of H$\alpha$ emission is greater or smaller than some threshold
value, typically 5--10 Angstroms. Of course, this boundary line is
somewhat arbitrary, and transitional cases are to be expected, but we
are aware of only one T~Tauri star, KH~15D, that seems to mock this
definition by alternating between classical and weak-lined with
clocklike regularity (Hamilton et al.\ 2003).

This is not the only unusual property of KH~15D. Although it was
classified as a variable member of the young cluster NGC~2264 by
Badalian \& Erastova (1970), the peculiarity of its variations was not
appreciated until Kearns \& Herbst (1998) found that the object fades
by 3~magnitudes every 48 days. Hamilton et al.\ (2001) and Herbst et
al.\ (2002) proposed that the star was being periodically eclipsed by
circumstellar material. Further monitoring revealed qualitative
changes in the eclipses, most notably a gradual increase in their
depth and duration, as well as the progressive disappearance of a
``re-brightening'' event that once characterized the mid-eclipse
phase. The dramatic growth in the equivalent width of H$\alpha$
emission occurs during the transition from the bright state to the
faint state. Apparently the occulting material blocks the photosphere
of the star and thereby reduces the contrast between the photosphere
and the H$\alpha$-emitting region: the star has a ``natural
coronagraph'' (Hamilton et al.\ 2003). Eclipse-related variations in
both optical polarization (Agol et al.\ 2004) and molecular H$_2$
emission (Deming, Charbonneau, \& Harrington 2004) have also been
observed.

For several years, the cause of the eclipses and of their evolution
were the subject of much speculation (Grinin \& Tambovtseva 2002,
Herbst et al.\ 2002, Winn et al.\ 2003, Barge \& Viton 2003, Klahr \&
Bodenheimer 2003, Agol et al.\ 2004, Barrado y Navascu{\'e}s et al.\
2004). Recently it was proposed that KH~15D is an eccentric binary
system that is being occulted by the edge of a circumbinary disk (Winn
et al.\ 2004, Chiang \& Murray-Clay 2004). In this scenario, we are
viewing the system nearly edge-on, and the sky projection of the disk
acts as a dark screen that gradually covers the binary. The advance of
the screen is a consequence of the nodal precession of the disk, which
is inclined relative to the binary plane. At present, one member of
the binary is hidden at all orbital phases, and the other member
undergoes an eclipse each time its orbital motion brings it behind the
screen. Winn et al.\ (2004; hereafter, Paper~I) showed that this model
unified the diverse properties of KH~15D and provided a quantitative
fit to the available measurements of the eclipses and their
evolution. Chiang \& Murray-Clay (2004) arrived at the same idea and
argued that the disk must be warped and radially narrow (a ``ring'')
in order to maintain rigid precession.

It has since been confirmed that KH~15D is a spectroscopic binary with
an eccentric orbit (Johnson et al.\ 2004). In addition, a wealth of
new photometric data has become available, from analyses of archival
photographs (Johnson et al.\ 2005, Maffei, Ciprini, \& Tosti 2005),
and modern CCD observations (Hamilton et al.\ 2005; Barsunova, Grinin,
\& Sergeev 2005; Kusakabe et al.\ 2005). Our main motivation for the
work described in this paper was to test whether or not the
circumbinary-disk model is consistent with these new data. In \S\S~2
and 3, we describe our compilation of the data, and in \S~4 we specify
the model and its parameters.  Section 5 is a direct extension of
Paper~I: we attempt to fit the observed times and durations of the
eclipses, and the observed radial-velocity variations.

A second motivation for this work was to understand certain features
of the data that had not previously been explained. What is the origin
of the light that is received during eclipses? Why is the depth of the
eclipses slowly increasing with time? What causes the
nearly-repeatable flux variations that are seen throughout ingress,
mid-eclipse, and egress?  We found that by making a minor elaboration
to the model---surrounding each of the stars with a faint blue
halo---all of these observations could be accurately described. This
is shown in \S~6.

Finally, in \S~7 we discuss the interpretation of the fitted model
parameters, in particular those that describe the halos. The most
likely interpretation is that scattering by the disk creates an
apparent halo around each star. We suggest some future observations
and calculations that may clarify the interpretation. If these are
successful, and if the occulting edge is as sharp as we suspect, then
it will be possible to use the occulting edge to map out the
environment of the visible star with wondrously fine detail. This
would broaden the significance of this system from being a fascinating
puzzle to being a crucial fortuitous case that will provide a deeper
understanding of all young stars.

\section{Photometry}
\label{sec:photometry}

Two types of photometry are available for this system: photographic
photometry, generally from years before 1985, and CCD photometry, from
1995 to the present. In all cases, the measurements refer to the total
light from the system, because the system has not been spatially
resolved (with the possible exception of an H$_2$ filament observed by
Tokunaga et al.\ 2004). The data are very heterogeneous, having been
gathered with dozens of different telescopes and reduced with
different procedures. This section describes the available data and
our efforts to allow them to be modeled within a single framework. In
general, our attitude was cautious, in the sense that we wished to
maximize the reliability of the data even at the expense of discarding
some of the data.

\subsection{Photographic Photometry}
\label{subsec:plates}

As a fairly unobscured example of a young stellar cluster, rich with
variable stars, NGC~2264 has long been a popular target for
photometric campaigns. Many photographic plates of NGC~2264 taken over
the past century have been preserved in observatory archives. Four
analyses of archival observations of KH~15D have been published:

\begin{enumerate}

\item Winn et al.\ (2003) used blue-sensitive exposures from the
Harvard College Observatory to show that in the first half of the 20th
century, the system was rarely (if ever) fainter than its modern
out-of-eclipse level by more than 1~magnitude. Given the number of
plates analyzed, they concluded that the system spent less than 20\%
of the time in such a faint state. More accurate photometry was
impossible because of the glare from a nearby star, HD~47887.

\item Johnson \& Winn (2004) performed profile photometry on a
digitized set of infrared-sensitive exposures from the 92/67~cm
Schmidt telescope of the Astrophysical Observatory of Asiago, Italy,
from the time period 1967--1982. The glare from HD~47887 is less
problematic on the infrared plates than on the blue plates. The
relative magnitudes have a typical accuracy of 0.1~mag. They were
placed on the standard Cousins $I$ band magnitude scale using a set of
reference stars whose $I$ band magnitudes had been measured with CCD
observations by Flaccomio et al.\ (1999). The zero point uncertainty
is 0.14~mag.

\item Maffei et al.\ (2005) analyzed a collection of blue and infrared
plates that were originally obtained by the lead author, P.\ Maffei,
between 1955 and 1970 using 5 different telescopes. Relative
photometry was performed visually, with an estimated accuracy ranging
from 0.05~mag to 0.2~mag depending on the telescope. The blue
magnitudes were converted to the Johnson $B$ band using a set of
reference stars with photographic and photoelectric $B$ magnitudes
given by Walker (1956). The infrared magnitudes were converted to the
Cousins $I$ band, after adopting an $I$-band magnitude for each
reference star based on an extrapolation of its $B-V$ color.

\item Johnson et al.\ (2005) gathered and digitized a collection of 87
plates, representing 8 different telescopes and 22 different
filter/emulsion combinations, from the time period 1954--1997. The
magnitudes of KH~15D were measured and reduced to $UBVRI$ magnitudes
using the procedure of Johnson \& Winn (2004).

\end{enumerate}

\begin{figure}[!h]
\epsscale{1.2}
\plotone{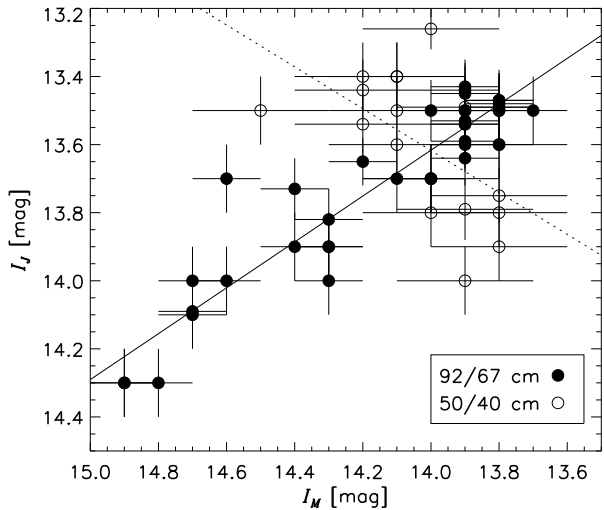}
\caption{ Comparison between the $I$-band magnitudes of KH~15D
reported by two different groups based on the same photographic
plates. Each point represents a single infrared plate, with $I_M$
defined as the value reported by Maffei et al.\ (2005), and $I_J$ as
the value reported by either Johnson \& Winn (2004) or Johnson et al.\
(2005).  Different symbols correspond to different telescopes.  The
solid line is a linear least-squares fit to the 92/67~cm data, and the
dotted line is a linear least-squares fit to the 50/40~cm data.
\vspace{0.2in}
\label{fig:maffei}}
\end{figure}

The latter 3 samples have 56 plates in common, giving us the
opportunity to check for consistency between the different photometric
methods. The plates in common are all infrared plates from one of the
two Asiago Schmidt telescopes: 38 are from the 92/67~cm telescope, and
18 are from the 50/40~cm telescope. Figure~\ref{fig:maffei} shows a
comparison of $I_M$, the magnitude reported by Maffei et al.\ (2005),
and $I_J$, the magnitude reported by either Johnson \& Winn (2004) or
Johnson et al.\ (2005) based on the same plate. There are significant
discrepancies. For the 92/67~cm plates, the relation between $I_J$ and
$I_M$ is well-fitted by a straight line,
\begin{equation}
I_J - 13.80 = 0.67 (I_M - 14.27),
\label{eq:maffei}
\end{equation}
as plotted in Fig.~\ref{fig:maffei}. This relation may reflect the
different methods for determining the zero point and for coping with
the nonlinear response of the emulsions. For the 50/40~cm plates,
there is no obvious relation between $I_J$ and $I_M$. The best-fitting
line (the dotted line in Fig.~\ref{fig:maffei}) represents an
$anti$-correlation between the two sets of results, which does not
make sense, and suggests that the photometric uncertainties have been
underestimated by one or both sets of authors. Of all the infrared
plates, Johnson et al.\ (2005) found the 50/40~cm plates to be the
most difficult to analyze, because of the large plate scale and
consequently poor sampling of the stellar images.

In this work, we attempt to fit only the $I$-band data, mainly because
the majority of modern CCD observations were made in the $I$ band, and
also because we judge the photographic photometry to be most reliable
for the infrared plates. We briefly discuss the data at other
wavelengths in \S~\ref{sec:discussion}, but we do not use those data
to determine our model parameters. Because we do not know how to
resolve the discrepancy between the two different sets of $I$-band
results, we do not consider the Asiago 50/40~cm data further. For the
Asiago 92/67~cm data, we use $I_J$ when it is available, and otherwise
we place $I_M$ onto the $I_J$ system\footnote{We transformed $I_M$
  into $I_J$, rather than the other way around, because the zero point
  of the $I_J$ scale was set with reference to CCD $I$-band
  observations, and we therefore expected $I_J$ to provide better
  consistency with the modern photometry.} using
Eq.~(\ref{eq:maffei}). For the data from all of the other telescopes,
we use the published results without modification. The resulting data
set consists of 81 measurements. The data are plotted in
Fig.~\ref{fig:plotall} along with the CCD photometry from more recent
times.

\subsection{CCD Photometry}
\label{subsec:ccd}

Over the last decade, KH~15D has been monitored intensively. The first
few years of observations used the 0.6~m telescope at Van Vleck
Observatory on the campus of Wesleyan University, in Connecticut, and
were designed to monitor the variability of a large sample of stars in
NGC~2264. After two seasons of observing, Kearns \& Herbst (1998)
identified KH~15D as an especially interesting variable star. Hamilton
et al.\ (2001) focused specifically on KH~15D, presenting a third year
of photometry along with optical spectroscopy and a discussion of
possible explanations for the peculiar eclipses. An observing campaign
involving many observatories was organized, and the first year's
results were described by Herbst et al.\ (2002). Most recently,
Hamilton et al.\ (2005) presented both new photometry and a
comprehensive summary of 9 years of CCD photometry involving a dozen
different telescopes. We refer the reader to that work for a detailed
discussion of the calibration and internal consistency of the various
data sets involved. Most of the photometry is in the Cousins $I$
band. A notable exception is the season of 2001--2002, when
simultaneous Johnson $VI$ measurements were made with the Yale 1~m
telescope on nearly every clear night. There are also a smaller number
of other observations in the $UBVR$ bands.

Barsunova et al.\ (2005) have independently monitored KH~15D in the
Johnson $V$ and Cousins $R$ and $I$ bands since 2002, using the 0.7~m
telescope of the Crimean Astrophysical Observatory. Eighteen of their
$I$ band measurements were taken within 6 hours of a measurement by
Hamilton et al.\ (2005), which, as in the previous section, gives us
an opportunity to check for consistency between the teams' different
photometric procedures. We found them to be consistent with the linear
transformation
\begin{equation}
I_H = 0.97 I_B + 0.27,
\label{eq:grinin}
\end{equation}
which is essentially a difference in the zero point of the magnitude
scale.

Recently, Kusakabe et al.\ (2005) presented near-infrared ($JHK_s$)
photometry of KH~15D spanning two observing seasons, from 2003 to
2005. As stated previously, in what follows we are concerned almost
entirely with the $I$-band data, although we do remark on the
near-infrared data in \S~\ref{sec:discussion}.

For our model-fitting purposes, we considered all of the $I$-band data
of Hamilton et al.\ (2005) and Barsunova et al.\ (2005), after
transforming the latter data set according to Eq.~(\ref{eq:grinin})
for the sake of uniformity. In addition, V.\ Grinin kindly provided us
the results from the 2004--2005 observing season that have not yet
been published, which we also transformed according to
Eq.~(\ref{eq:grinin}) and added to the data set under
consideration. In total, there are 6780 data points, of which 87 are
from the Russian group. The uncertainties range widely from a few
millimagnitudes, for some observations when KH~15D was in its bright
state, to 0.25~mag, when it was in its faint state.

\subsection{Summary of the Photometry}
\label{subsec:summary}

\begin{figure*}[t]
\begin{center}
  \leavevmode
\hbox{%
  \epsfxsize=6in
  \epsffile{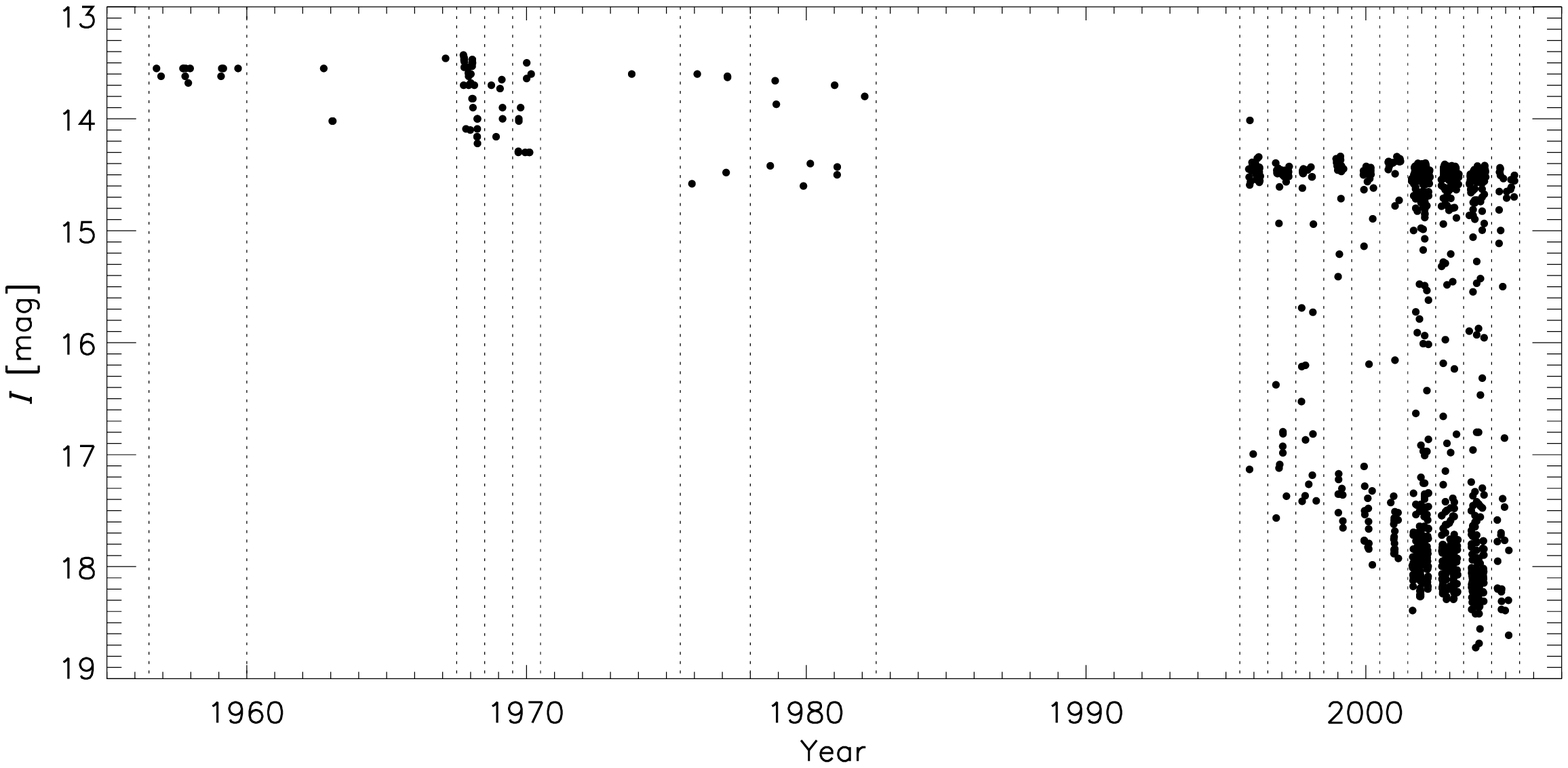}}
\end{center}
\caption{\label{fig:plotall}
Fifty years of Cousins $I$-band photometry of KH~15D. The
post-1995 data have been averaged into 6~hour bins. (The dotted lines
mark the time ranges that are plotted in the panels of
Fig.~\ref{fig:bestfit3_lc}.
\vspace{0.2in}
}
\end{figure*}

The entire $I$-band time series, from 1956 until the present, is shown
in Fig.~\ref{fig:plotall}. Many interesting features of the system's
photometric history can be seen in this figure. Before about 1960,
there is not much evidence for variability greater than
0.1~mag.\footnote{Maffei et al.\ (2005) found evidence for periodic
variations in $B$ between 1958 and 1963, a time when $I$-band
variations at the same level could be excluded. Most of the relevant
data is from the Asiago 50/40~cm Schmidt telescope. In
\S~\ref{subsec:plates} we showed that two different groups arrived at
significantly different results when analyzing the same infrared
plates from that telescope, suggesting that the uncertainties may be
larger than they appear. The uncertainties should be still larger for
the blue plates because of the difficulty presented by the neighboring
bright star. Hence, although the possibility that the eclipses began
in $B$ prior to $I$ is interesting and worthy of further
investigation, we do not consider this possibility further in this
paper.} Beginning in the early 1960s and lasting until at least the
early 1980s, the system varied between $I\approx 13.6$ and 14.5. By
1995, the variations were between 14.5 and a faint state that has
gradually darkened from about 17 to 18.5. Although it is not apparent
in Fig.~\ref{fig:plotall}, both the pre-1980 and post-1995 variations
are periodic with a 48.4~day period, and the phase of minimum light
shifted by nearly 180~degrees between those time periods (Johnson \&
Winn 2004).

\section{Radial Velocities}
\label{sec:rv}

Possible time variations of the Doppler shift of starlight from KH~15D
were first reported by Hamilton et al.\ (2001) and Hamilton et al.\
(2003). Subsequently, Johnson et al.\ (2004) showed that KH~15D is
indeed a spectroscopic binary, based on a series of high-resolution
optical spectra using telescopes at the Keck, Magellan, and McDonald
observatories. With 16 measurements spanning 1.3~yr with accuracies
ranging from 0.2 to 0.6~km~s$^{-1}$, Johnson et al.\ (2004) showed
that the Doppler shifts could be explained as radial velocity
variations of a star in an eccentric Keplerian orbit.

\begin{figure}[!h]
\epsscale{1.25}
\plotone{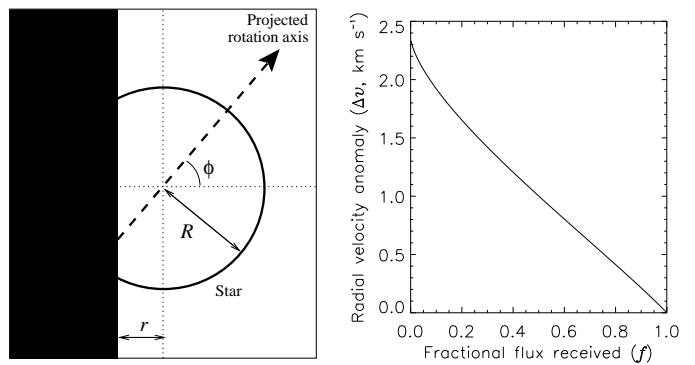}
\caption{ The Rossiter-McLaughlin effect produced by a straight-edged,
semi-infinite occultor. {\it Left.}---Diagram of the star, its
projected rotation axis, and the occulting edge. {\it Right.}---The
magnitude of the Rossiter-McLaughlin effect, using
Eqs.~(\ref{eq:ross1}) and (\ref{eq:ross2}) and assuming $V_{\rm rot}
\sin I_\star = 6.9$~km~s$^{-1}$ and $\phi = 20\arcdeg$, as estimated
for KH~15D.
\label{fig:ross}}
\end{figure}

If the occulting edge of KH~15D is sharp enough to resolve the stellar
photosphere, then the radial velocity measurements taken during
ingress or egress are subject to systematic errors due to the
Rossiter-McLaughlin effect (Rossiter 1924, McLaughlin 1924). This term
refers to the apparent velocity perturbation of a partially occulted
star due to stellar rotation. For example, if only the receding half
of a star is visible, then the starlight is redshifted in excess of
the shift one would expect from only the star's center-of-mass
motion. This effect has long been observed during eclipses of binary
stars (see, e.g., Worek 1996, Rauch \& Werner 2003), and more recently
during transits of extrasolar planets (Bundy \& Marcy 2000, Queloz et
al.\ 2000, Snellen 2005, Winn et al.\ 2005). To estimate the magnitude
of the effect, suppose that a star of radius $R$ is occulted by a
straight-edged semi-infinite screen, as depicted in the left panel of
Fig.~\ref{fig:ross}. The minimum distance between the edge and the
stellar center is $r$. The angle between the normal to the edge and
the sky projection of the stellar north pole is $\phi$. The fraction
of the total stellar flux ($f$) and the mean line-of-sight velocity
($\Delta V$) of the exposed portion of the star are given by
\begin{eqnarray}
f        & = & \frac{1}{\pi} \left( \cos^{-1} u - u\sqrt{1-u^2} \right),
\label{eq:ross1} \\
\Delta V & = & \frac{2}{3} \hskip 0.05in V_{\rm rot} \sin I_\star \sin\phi
               \left[
                 \frac{ (1-u^2)^{3/2} }{ \cos^{-1}u - u\sqrt{1-u^2} }
               \right],
\label{eq:ross2}
\end{eqnarray}
where $u = r/R$, $V_{\rm rot}$ is the stellar rotation speed and
$I_\star$ is the inclination of the stellar rotation axis with respect
to the line of sight. These expressions neglect differential rotation
and limb darkening.

For the case of KH~15D, Hamilton et al.\ (2005) measured $V_{\rm rot}
\sin I_\star = 6.9\pm 0.3$~km~s$^{-1}$ for the currently visible
star. We estimate $\phi\approx 20\arcdeg$, using the model of Paper~I,
and assuming the stellar rotation axis to be perpendicular to the
stellar orbital plane. The right panel of Fig.~\ref{fig:ross} shows
the resulting magnitude of $\Delta V$ as a function of $f$. For the
perturbations to be smaller than 0.2~km~s$^{-1}$, and hence smaller
than the measurement uncertainties in the radial velocity data, we
need $f>0.9$. Thus, for model-fitting purposes, we consider only the
12 radial velocity measurements that were obtained when the system
flux was 90\% or greater of its mean out-of-eclipse flux. This is a
cautious procedure, in the sense that the systematic error is smaller
for the more realistic case of a star with limb darkening and an
occulting edge that is not perfectly sharp.\footnote{We also tried
  using all the data with $f>0$, after enlarging the error bars to
  encompass the expected systematic errors. In that case, the
  procedure described in \S~\ref{sec:model} resulted in significantly
  poorer fits, but the optimized values of the parameters were similar
  to those reported in Table~1.}

We made two other minor adjustments to the data published by Johnson
et al.\ (2004). First, we corrected the time stamps on some of the
Keck measurements, which were in error by 0.5~day because of a
confusion between Julian and Modified Julian dates. Second, we added
0.25~km~s$^{-1}$ in quadrature to the uncertainties of the data
labeled ``b'' in Table~1 of Johnson et al.\ (2004). In those cases,
the radial velocity scale was calibrated using only one reference star
and the quoted error estimate was statistical, i.e., based only on the
noise in the spectrum of KH~15D. In the other cases (labeled ``a''),
multiple reference stars were available, and it was found that the
statistical error was smaller than the variance in the results of
using different choices for the reference star. Using the ``a''
results we estimated that the systematic error in the calibration is
$\approx$0.25~km~s$^{-1}$. For convenience, the complete and updated
list of radial velocity measurements is given in
Table~\ref{tbl:rv}. Only the 12 measurements with $f\geq 0.9$ were
used in the model-fitting procedure.

\begin{deluxetable}{ccc}
\tabletypesize{\small}
\tablecaption{Radial Velocity Measurements of KH~15D}
\tablewidth{0pt}
\tablehead{
\colhead{Julian Date} &
\colhead{Relative flux ($f$)} &
\colhead{$v_r$}
}
\startdata
$2452242.7649$ & 0.0 & $ 9.00 \pm  0.32$ \\
$2452263.7640$ & 0.0 & $12.30 \pm  0.60$ \\
$2452573.0749$ & 1.0 & $ 1.70 \pm  0.20$ \\
$2452576.0111$ & 1.0 & $ 3.00 \pm  0.30$ \\
$2452624.7388$ & 1.0 & $ 3.10 \pm  0.32$ \\
$2452653.9689$ & 0.6 & $ 3.30 \pm  0.20$ \\
$2452678.9007$ & 0.7 & $ 9.00 \pm  0.56$ \\
$2452679.9109$ & 0.2 & $11.50 \pm  0.56$ \\
$2452707.5117$ & 1.0 & $ 0.80 \pm  0.32$ \\
$2452947.0845$ & 0.9 & $ 1.20 \pm  0.47$ \\
$2452948.0948$ & 0.9 & $ 1.80 \pm  0.39$ \\
$2453008.7620$ & 1.0 & $ 1.40 \pm  0.65$ \\
$2453009.8520$ & 1.0 & $ 1.70 \pm  0.47$ \\
$2453014.7600$ & 0.9 & $ 5.80 \pm  0.47$ \\
$2453014.8208$ & 0.9 & $ 5.40 \pm  0.30$ \\
$2453016.0063$ & 0.4 & $ 7.00 \pm  0.32$ \\
$2453044.8341$ & 1.0 & $ 1.80 \pm  0.20$ \\
$2453045.8282$ & 1.0 & $ 1.25 \pm  0.20$
\enddata
\tablecomments{ From Hamilton et al.\ (2003) and Johnson et al.\
(2004), modified as described in \S~\ref{sec:rv}. Positive velocity
corresponds to motion away from the Sun. }
\label{tbl:rv}
\end{deluxetable}

\section{The Model of KH~15D}
\label{sec:model}

\subsection{Definition of Parameters}

Our model consists of two stars (A and B) in a Keplerian orbit, and a
straight-edged semi-infinite occulting screen representing the sky
projection of a circumbinary disk. Star A is the star that is
currently undergoing periodic eclipses, and for which radial velocity
variations have been measured. The stars have masses denoted by $M_A$
and $M_B$, and radii $R_A$ and $R_B$. The orbit is specified by the
period ($P$), eccentricity ($e$), argument of pericenter ($\omega$),
inclination with respect to the sky plane ($I$), heliocentric radial
velocity of the center of mass ($\gamma$), and a particular time of
pericenter passage ($T_p$).

\begin{figure}[!h]
\epsscale{1.0}
\plotone{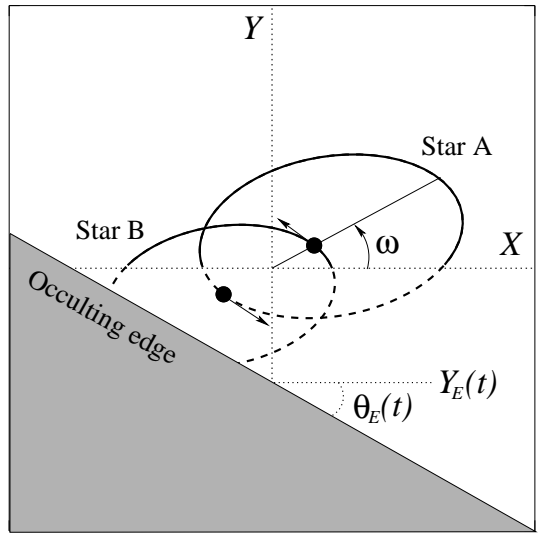}
\caption{ The coordinate system. The $X$--$Y$ plane is the sky plane
and the $Z$-axis points towards the observer.  The orbital plane is
inclined by an angle $I$ with respect to the sky plane.  The line of
nodes is the $X$-axis, with the ascending node of star B occurring at
$X>0$.  The $Z>0$ portions of the orbit are marked with solid lines,
and the $Z<0$ portions are marked with dashed lines.  The stars are
shown at pericenter, and $\omega$ is the argument of pericenter.  The
instantaneous location of the occulting edge is specified by $Y_E$,
its intersection with the $Y$-axis, and by $\theta_E$, the angle that
the edge makes with the $X$-axis.  A configuration with $I<90\arcdeg$,
$Y_E<0$, and $\theta_E<0$ is shown.
\label{fig:coord_diagram}}
\end{figure}

A diagram of the coordinate system\footnote{Although our model is the
  same as that presented previously in Paper~I, we have chosen a
  different coordinate system for convenience.} is given in
Fig.~\ref{fig:coord_diagram}. The $X$--$Y$ plane is the sky plane. The
$X$-axis is chosen to be along the line of nodes, with the ascending
node of star B at $X>0$. The location and orientation of the occulting
screen are specified by its point of intersection between the edge and
the $Y$-axis ($Y_E$), and by the angle that the edge makes with the
$X$-axis ($\theta_E$).  In Paper~I, we assumed that the screen moves
uniformly and without rotation, i.e., $Y_E$ was taken to be a linear
function of time, and $\theta_E$ was taken to be a constant. In
\S~\ref{sec:orbit} we consider these same assumptions, but we also
consider more general cases that we argue are more realistic.

\begin{figure}[!h]
\epsscale{1.0}
\plotone{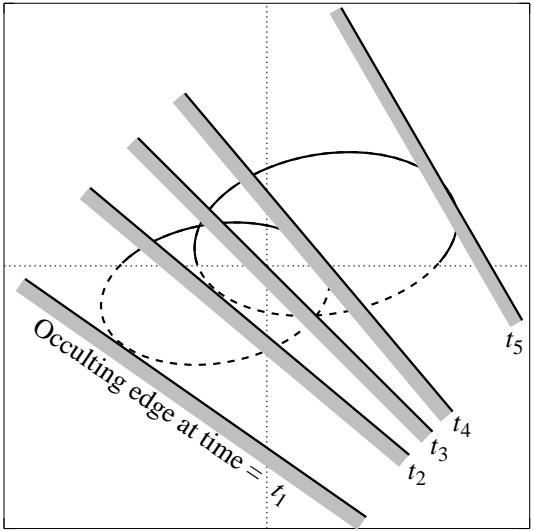}
\caption{ Definition of the 5 orbital contact times.  In this
illustration, the occulting screen has $\theta_E(t_3) < 0$,
$\dot{\theta_E} > 0$, and $\dot{Y_E} > 0$.
\label{fig:contact_times}}
\end{figure}

Once the orbital parameters and the screen's trajectory are specified,
five ``orbital contact times'' can be defined (see
Fig.~\ref{fig:contact_times}). First contact ($t_1$) occurs when the
occulting edge is tangent to the sky-projected orbit of star B for the
first time. Before first contact, no eclipses
occur. Fig.~\ref{fig:plotall} suggests that first contact occurred
between 1959 and 1967. Second contact ($t_2$) occurs when the
occulting edge is first tangent to the sky-projected orbit of star A.
Between $t_1$ and $t_2$, star B is periodically occulted, but we
receive constant light from star A. The total light exhibits
``partial'' or ``diluted'' eclipses, as were observed between the late
1960s and the early 1980s. After $t_2$, both stars undergo periodic
eclipses, and the light curve can appear quite complex. Third contact
($t_3$) occurs when the edge crosses the projected center of mass of
the binary.  There are no abrupt changes in the photometric
behavior. After $t_3$, if the eccentricity is large, star B is seen
for a rapidly shrinking fraction of the orbital period; it appears
only briefly during the middle of each eclipse of star A. This is the
era of ``central re-brightenings,'' as observed in 1996 and 1997.
After fourth contact ($t_4$), the entire orbit of star B is hidden. As
the screen advances further, the duration of the eclipses of star A
increases until fifth contact ($t_5$) when the entire system is
covered.

\subsection{Interpretation of the Occulting Screen}
\label{subsec:screen}

Of course, the real occulting edge is not semi-infinite. The
dimensions of the screen's true sky projection depend on the
three-dimensional structure of the circumbinary disk. Star B should
eventually reappear, possibly even before $t_5$, and the qualitative
changes in total-light photometry that have been observed since 1960
may occur in reverse order as the trailing edge of the screen uncovers
the binary. Those phenomena are not included in the model because the
true three-dimensional disk structure is not known and cannot be
determined from existing observations. However, some remarks are in
order regarding the relation between the idealized screen and a real
circumbinary disk.

\begin{figure*}[t]
\begin{center}
  \leavevmode
\hbox{%
  \epsfxsize=6in
  \epsffile{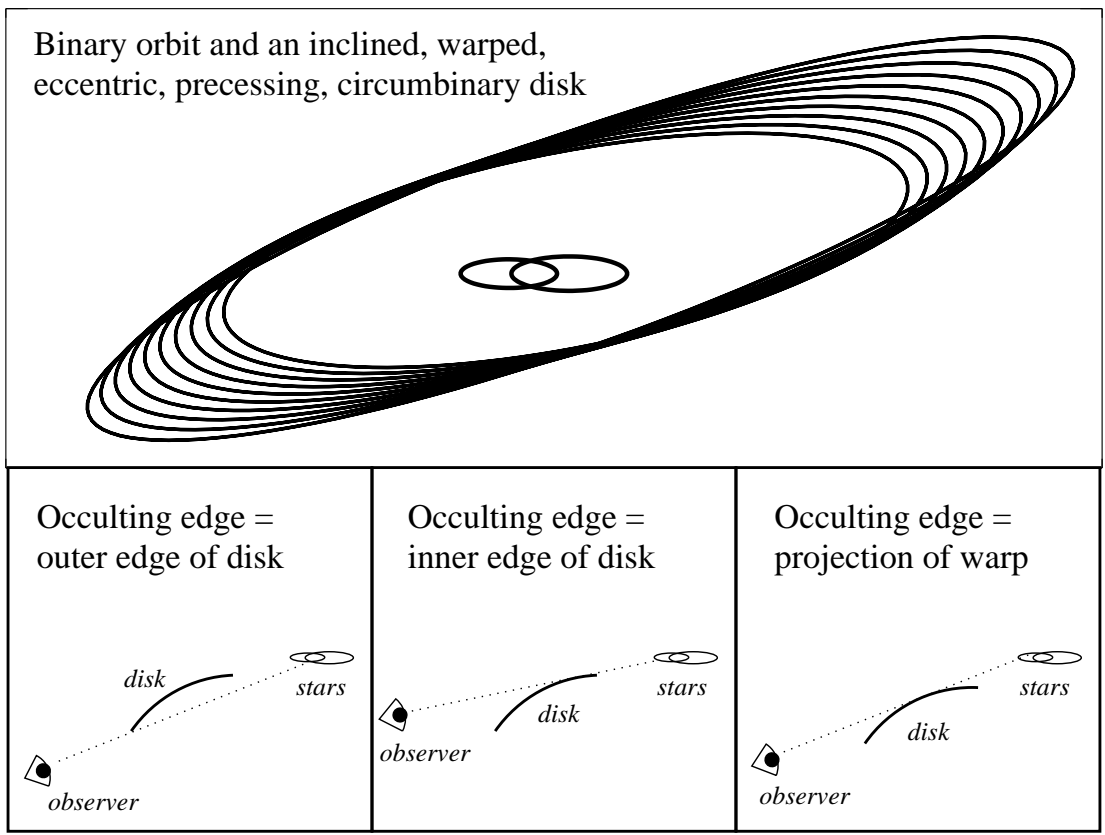}}
\end{center}
\caption{\label{fig:diskviews}
An illustration of the binary and the surrounding disk
(above), and some possible viewing geometries (below).  The true
three-dimensional disk structure is not known.  In particular the warp
may have the opposite sign of the inclination gradient ($dI_r/da_r$)
as the warp depicted.  In the lower panels, for clarity, only the
cross-section of the disk in the sky plane is shown.
\vspace{0.4in}
}
\end{figure*}

The premise of the theory is that the circumbinary disk is inclined
with respect to the binary plane, which causes the orbits of the
disk's constituent particles to precess. Chiang \& Murray-Clay (2004)
showed that the disk should be warped, and that its inner and outer
radii should be of the same order of magnitude, in order for the
``ring'' to precess as a unit. The edge of the screen may represent
the ring's inner or outer edge in projection, or it may be a
projection of the warp, in which case the edge does not necessarily
represent a single orbital distance. These possibilities are depicted
in Fig.~\ref{fig:diskviews}.

The correspondence between the precession of the disk and the motion
of the screen is harder to visualize. For simplicity, imagine that the
binary is surrounded by a circular ring of radius $a_r$ and
inclination $I_r$ with respect to the binary's orbital plane. The ring
represents, say, the inner radius of the disk, and its sky projection
is the occulting feature. The ring's line of nodes (LON) precesses
around the orbital plane with an angular frequency $\beta <
0$.\footnote{The negative value of $\beta$ means that the LON
regresses relative to the orbital angular momentum vector.} Now
imagine viewing the stellar orbits edge-on. When the LON of the ring
is pointed at the observer, the ring appears as a straight line with
$Y_E=0$ and $\theta_E=I_r$. This is $t_3$, the moment of third
contact. At other times, the sky projection of the ring is an
ellipse. Considering the $Z>0$ half of the ellipse (i.e.\ the portion
that is in front of the binary), the intersection point with the
$Y$-axis and the angle it makes with the $X$-axis are
\begin{eqnarray}
Y_E(t) & = &
       a_r \hskip 0.05in
        \left[
           \frac{\tan I_r \sin \beta (t-t_3)}
                {\sqrt{1 + \sin^2{\beta (t-t_3)} \tan^2 I_r}}
        \right], \\
\theta_E(t) & = &
       \tan^{-1}
          \left[
             \frac{\tan I_r \cos\beta(t-t_3) \left[\sin\beta(t-t_3) \cos I \tan I_r + \sin I\right] }
                  {\sqrt{1 + \cos^2\beta(t-t_3) \tan^2 \beta(t-t_3)}}
          \right].
\end{eqnarray}
For small $I_r$, these expressions reduce to $Y_E(t) = a_r \tan I_r
\sin\beta(t-t_3)$ and $\theta_E(t) = I_r \cos\beta(t-t_3)$. For times
close to $t_3$, $\dot{Y}_E(t) = \beta a_r \tan I_r =$~constant, and
$\theta_E(t) = I_r = $~constant, in accordance with the assumptions
made in Paper~I.

However, the applicability of this approximation is not guaranteed.
Over a 50-year time span, the higher-order terms in $\beta(t-t_3)$ may
be appreciable. More importantly, the disk is probably elliptical,
given that the stellar binary has $e\sim 0.6$ (see Paper~I, Hamilton
et al.\ 2005, Herbst \& Moran 2005, and also \S~\ref{sec:orbit}
below). This alters both the amplitudes and phases of functions
$Y_E(t)$ and $\theta_E(t)$, depending on the eccentricity and argument
of pericenter of the ring. These functions may also depend on the
details of the warp. In addition, if $a_r$ is small enough, the
curvature of the occulting edge in the sky plane may be
significant. We found that there are too many possibilities for the
data to distinguish among them. Instead, our approach was to begin by
pursuing the simplest possibility, a straight-edged screen with a
constant angle and speed, while also preparing to allow for angular
rotation speeds (of order $\sim$$\beta I_r$, according to the formula
for a circular ring) and accelerations ($\sim$$\beta^2 a_r \tan I_r$)
if they appear to be needed.

\subsection{Expectations for Parameter Values}
\label{subsec:expectations}

Before proceeding to the model-fitting results, we review what was
already known of the parameter values from other observations or
theoretical arguments. Star A has been spectrally classified as K6-7,
with an effective temperature of approximately 4000~K and luminosity
0.4~$L_\odot$ (Hamilton et al.\ 2001, Agol et al.\ 2004), assuming a
cluster distance of 760~pc (Sung et al.\ 1997). Its mass was estimated
to be 0.6~$M_\odot$ by Flaccomio et al.\ (1999) and 0.5--1.0~$M_\odot$
by Hamilton et al.\ (2001), based on theoretical pre--main-sequence
(PMS) evolutionary tracks. We find the range $M_A = 0.6\pm
0.1~M_\odot$ to be in reasonable agreement with the six different sets
of PMS tracks compiled by Hillenbrand \& White (2004). The
corresponding radius estimate is $R_A = 1.3\pm 0.1~R_\odot$.
Unfortunately, we cannot estimate $M_A$ and $R_A$ independently using
our model and the current data, because the limited radial velocity
data place only weak bounds on the mass function of the binary, and
because there is a degeneracy between the stellar radius and the
projected orbital speed during occultations.

No spectrum of star B has been obtained, but its luminosity can be
estimated using simple considerations of the system's photometric
history. The re-brightenings observed between 1995 and 1998 (when star
B was seen alone) had a maximum flux of 1.4 times the flux of the
out-of-eclipse signal (when star A was seen alone), indicating that
$L_B/L_A\approx 1.4$. The photographic photometry is not as accurate,
but it also suggests $L_B/L_A \gsim 1$. The flux of the maximum-light
phase of the pre-1980 photometry (when both stars were seen) was
approximately double the flux of the minimum-light phase (when star A
was seen alone), and was approximately 2.3 times that of the modern
maximum-light phase. The PMS tracks of D'Antona \& Mazzitelli (1997)
predict a mass-luminosity relation $L \propto M^{1.9}$, and a
mass-radius relation $R \propto M^{0.27}$, leading to the expectations
$M_B/M_A \approx 1.2$ and $R_B/R_A \approx 1.05$.

Johnson et al.\ (2004) analyzed the radial velocity data alone. Given
the necessarily limited phase coverage of the observations, there were
strong degeneracies between many of the orbital parameters. Among the
most robust conclusions were that the orbital period is $P=48.4$~days,
the eccentricity is appreciable ($e > 0.3$), and the argument of
pericenter is within about 20$\arcdeg$ of zero. The period
determination is nearly identical with those based on periodograms of
the photometry (by, e.g., Hamilton et al.\ 2005, Johnson \& Winn 2004,
Maffei et al.\ 2005), although we note that the periodogram-based
method may be biased because the light curve is not strictly periodic.

Herbst \& Moran (2005) and Hamilton et al.\ (2005) made a theoretical
prediction for the eccentricity, based on the theory of tidal
interactions in close binaries. Tides will ``pseudo-synchronize'' the
stellar spins and orbit, meaning that the stellar angular-rotation
frequency, $\Omega_r$, and the orbital angular frequency at
pericenter, $n_p$, will reach a theoretically determined equilibrium
(Hut 1981):
\begin{eqnarray}
\frac{\Omega_r}{n_p} & = & \frac
{1 + \frac{15}{2} e^2 + \frac{45}{8} e^4 + \frac{5}{16} e^6}
{(1 + 3 e^2 + \frac{3}{8} e^4)(1 + e)^2} \nonumber \\
 & = & \left( \frac{P_{\rm orb}}{P_{\rm rot}} \right)
       \frac{(1-e^2)^{3/2}}{(1+e)^2}.
\end{eqnarray}
The time scale for this process is poorly known but is estimated to be
a few million years, which is approximately the age of the system. It
therefore seems likely that pseudo-synchronization is well underway,
even if it has not yet been achieved. Hamilton et al.\ (2005) measured
a rotation period $P_{\rm rot} = 9.6$~days for star A, from periodic
photometric variations observed out of eclipse and attributed to star
spots. Given that $P_{\rm orb} = 48.4$~days, the
pseudo-synchronization condition is $e=0.65\pm 0.01$. It seems likely
that the rotation of star A has been slowed down, rather than sped up,
given that its period is longer than usual for a low-mass star in
NGC~2264. If pseudo-synchronization has not yet been achieved, then
the equilibrium period must be even longer, and the eccentricity must
be correspondingly smaller to provide a slower angular velocity at
pericenter. Hence, whether or not pseudo-synchronization has been
achieved, the upper limit on the eccentricity is 0.66. This assumes
Hut's theory to be applicable, i.e., the effects of dynamical tides,
and of forces from the circumbinary disk, accreting material, or other
bodies in the system, can be neglected.

Some circumstantial evidence suggests that the orbital inclination is
nearly 90 degrees. First, the mere observation of eclipses suggests
that the system is viewed nearly edge-on, unless the circumbinary disk
is grossly misaligned with the orbital plane. Second, the inclination
of star A's rotation axis ($I_\star$) seems to be near 90 degrees.
The combination of the measured rotation period ($P_{\rm rot} =
9.6$~days), the projected rotation speed of the photosphere ($V_{\rm
rot} \sin I_\star = 6.9\pm 0.3$~km~s$^{-1}$), and the estimated
stellar radius ($R_A = 1.3\pm 0.1$~$R_\odot$) gives $\sin I_\star =
1.01 \pm 0.09$, or $I_\star > 67\arcdeg$. Third, Hamilton et al.\
(2003) observed a double-peaked emission line of [\ion{O}{1}], with
the peaks at radial velocities $\pm20$~km~s$^{-1}$ of the systemic
velocity. These are thought to arise from bipolar jets with space
velocities that are typically $\pm 200$~km~s$^{-1}$, suggesting that
the jets (and by implication the stellar poles and the orbit) are
inclined by $\gsim 80\arcdeg$.

Finally, an expectation for $\gamma$ (the heliocentric radial velocity
of the center of mass) is available thanks to the radial-velocity
survey of NGC~2264 by Soderblom et al.\ (1999). These authors found a
median radial velocity of $20\pm 3$~km~s$^{-1}$ among probable cluster
members. Thus, although Johnson et al.\ (2004) found the radial
velocity measurements to be compatible with any value of $\gamma$
between 7 and 22~km~s$^{-1}$, we expect the true value of $\gamma$ to
be near the high end of that range, assuming KH~15D to be a cluster
member.

\section{Models of the orbit}
\label{sec:orbit}

We take a two-step approach to the quantitative modeling of KH~15D.
The first step is a direct continuation of Paper~I. Rather than
fitting the photometry directly, we attempt to match some of its most
important properties: the eclipse durations and the ingress and egress
durations. For the moment, we ignore all of the other information in
the photometry, such as the light observed during eclipses, and the
detailed shapes of the ingress and egress light curves. The second
step, described in \S~\ref{sec:occultations}, is an attempt to fit the
full photometric time series.

To measure the eclipse durations, we plotted the phased light curve
for each year in which fairly complete phase coverage was achieved,
and determined the minimum and maximum phases of the half-flux points
of the eclipses. The results are given in Table~\ref{tbl:ecl}. They
agree with those given in Paper~I, and include new measurements based
on the newly available photometry, as well as a few measurements of
the eclipse fraction of star B based on the observations of central
re-brightenings (or lack thereof).

\begin{deluxetable}{cccc}
\tabletypesize{\small}
\tablecaption{Observed Bounds on Eclipse Fractions}
\tablewidth{0pt}
\tablehead{
\colhead{Star} &
\colhead{Year} &
\colhead{Lower Bound} &
\colhead{Upper Bound}
}
\startdata
B & 1958 & 0.000 & 0.050 \\
B & 1968 & 0.282 & 0.548 \\
B & 1970 & 0.211 & 0.634 \\
B & 1981 & 0.275 & 0.845 \\
B & 1996 & 0.734 & 0.957 \\
B & 1997 & 0.920 & 0.989 \\
A & 1997 & 0.252 & 0.360 \\
A & 1998 & 0.270 & 0.379 \\
A & 1999 & 0.303 & 0.569 \\
A & 2000 & 0.330 & 0.482 \\
B & 2000 & 1.000 & 1.000 \\
A & 2001 & 0.326 & 0.550 \\
A & 2002 & 0.410 & 0.432 \\
A & 2003 & 0.444 & 0.474 \\
A & 2004 & 0.497 & 0.539 \\
A & 2005 & 0.530 & 0.570
\enddata
\label{tbl:ecl}
\end{deluxetable}
\vspace{0.2in}

The ingress and egress durations are harder to measure because fine
time-sampling is needed. In Paper~I we estimated the durations based
on phased light curves from 2001--2003, but we see now that there is
too much variation between eclipses for this method to be
reliable. Instead, we used data from a single eclipse in 2002.1 that
was observed with especially high cadence. We considered only the
steepest (most rapidly varying) portion of the light curve between
flux levels 33\% and 67\% of the maximum, and fitted a model light
curve to these data consisting of a limb-darkened star being crossed
by a knife-edge occultor. The best-fitting ingress and egress
durations (the times taken to cross the diameter of the star) were
$2.55\pm 0.08$~days and $2.96\pm 0.05$~days, respectively.

We assume $M_A = 0.6~M_\odot$ and $R_A = 1.3~R_\odot$ as explained in
\S~\ref{subsec:expectations}. The mass of star B is taken to be a free
parameter. There are 6 free parameters describing the orbit: $\{P, e,
I, \omega, \gamma, T_p\}$. Additional free parameters are needed to
describe the occulting screen and its trajectory. We begin with the
case of a screen that moves uniformly with a fixed angle, in which
case 3 parameters are needed. One obvious parameterization, for
example, is $\{Y_E(T_p), \dot{Y_E}, \theta_E\}$. We use the
alternative parameter set $\{t_4, t_5, \theta_E\}$, which provides a
more direct connection to the observations.

The figure-of-merit function is
\begin{eqnarray}
\chi^2 & = & \chi^2_v + \chi^2_d \nonumber \\
       & = & \sum_{i=1}^{N_v} \left( \frac{ v_{C,i} - v_{O,i} } {\sigma_{v,i}} \right)^2 +
             \sum_{i=1}^{N_d} \left( \frac{ d_{C,i} - d_{O,i} } {\sigma_{d,i}} \right)^2,
\label{eq:chi2rv}
\end{eqnarray}
where $v_{O,i}$ are the observed radial velocities (of which there are
$N_v=12$), $\sigma_{v,i}$ are the corresponding 1~$\sigma$
uncertainties, and $v_{C,i}$ are the radial velocities as calculated
according to the model. The single measurement of the ingress duration
($d_{C,1}$) and the egress duration ($d_{C,2}$) are described with a
similar notation. We used an AMOEBA algorithm (Press et al.\ 1992, p.\
408) to minimize $\chi^2$, while simultaneously {\it requiring} the
calculated eclipse durations to obey the 16 constraints\footnote{The
eclipse duration measurements have the character of strict bounds,
rather than central values and 1~$\sigma$ uncertainties, because the
measurements are limited by time-sampling of the events rather than
statistical errors in the flux measurements.} given in
Table~\ref{tbl:ecl}. We also enforced the proper phase alignment
between the radial velocity data and the photometry by requiring the
model system to experience mid-eclipse within 5 days of
JD~2,452,352.5, a particular mid-eclipse observed in 2002.

\begin{figure*}[t]
\begin{center}
  \leavevmode
\hbox{%
  \epsfxsize=7in
  \epsffile{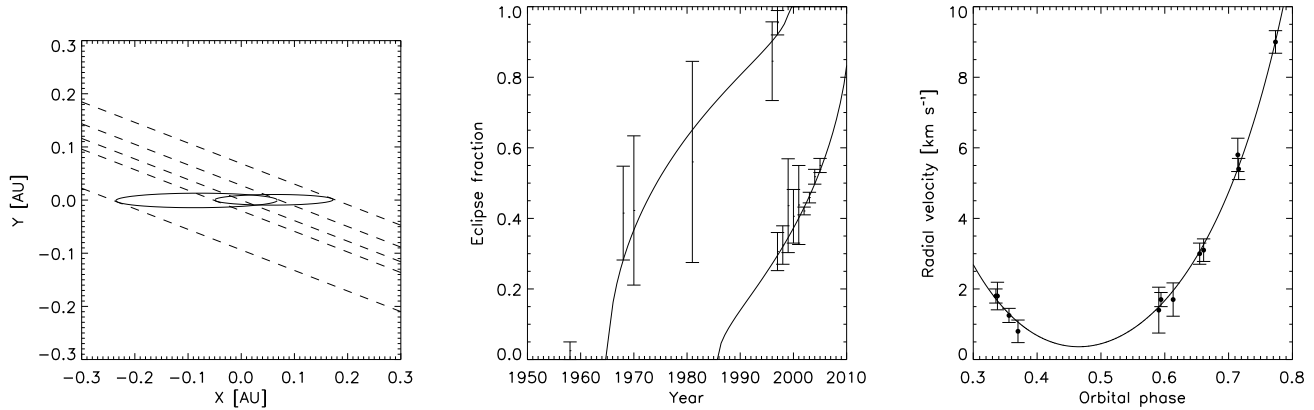}}
\end{center}
\caption{\label{fig:bestfit1}
{\bf Model 1}: $M_B$ is a free parameter and $\dot{\theta}_E
= 0$.  {\it Left.}---The binary orbit (solid lines) and the position
of the occulting edge at the 5 orbital contact times (dashed lines).
{\it Middle.}---The eclipse fractions of stars A and B as a function
of time, as observed (error bars) and calculated (solid lines).  {\it
Right.}---The radial velocity variations of star A as a function of
orbital phase, as observed (points with error bars) and calculated
(solid line).
}
\end{figure*}

An excellent fit is achieved, with $\chi^2 = 3$ and 4 degrees of
freedom. Thus the data are simultaneously consistent with the
measurements of radial velocity, ingress and egress duration, and
eclipse duration.  Figure~\ref{fig:bestfit1} shows the fit to the
radial-velocity and eclipse-fraction data. It also shows a diagram of
the orbit and the screen at each of the 5 orbital contact times. The
best-fitting parameter values are given in Table~\ref{tbl:params},
under the heading ``Model 1.''

\medskip

Many of the best-fitting parameter values conform with the
expectations given in \S~\ref{subsec:expectations}: the period is
$P=48.4$~days, the eccentricity is $e=0.55$, the inclination is
$83\arcdeg$, and the argument of pericenter is $6\arcdeg$.  However,
there are two exceptions. First, the heliocentric radial velocity of
the center of mass (14 km~s$^{-1}$) is somewhat low; KH~15D would be a
$\sim$2$~\sigma$ outlier in the radial velocity distribution of the
cluster. A second and much more serious problem is that the mass ratio
is $M_B/M_A = 0.76$, which is less than unity, even though it seems
certain that $L_B/L_A > 1$.

The latter problem can be traced to the assumption that the occulting
screen moves uniformly in a constant direction. Under this assumption,
$t_5-t_2$ is proportional to the projected size of star A's orbit,
which in turn is proportional to $M_A^{-1}$. Likewise, $t_4 - t_1
\propto M_B^{-1}$. The photometry provides the constraints $1959 < t_1
< 1967$, $1981 < t_2 < 1996$, $1997 < t_4 < 2000$, and $t_5 > 2005$,
and an extrapolation of the rapidly rising eclipse fraction suggests
$t_5 < 2010$. Together, these inequalities give
\begin{equation}
\frac{M_B}{M_A} = \frac{t_5 - t_2}{t_4 - t_1} < 0.97
\mbox{   (assuming constant screen velocity).}
\end{equation}
If $M_B/M_A$ is held fixed at 1.2, the value expected from a
consideration of the luminosity ratio, then it is impossible to fit
the eclipse data. For example, in the model, star B is covered 6 years
too early, or the rise in the eclipse fraction of star A over the last
decade is much slower than has been observed.

\begin{figure*}[t]
\begin{center}
  \leavevmode
\hbox{%
  \epsfxsize=7in
  \epsffile{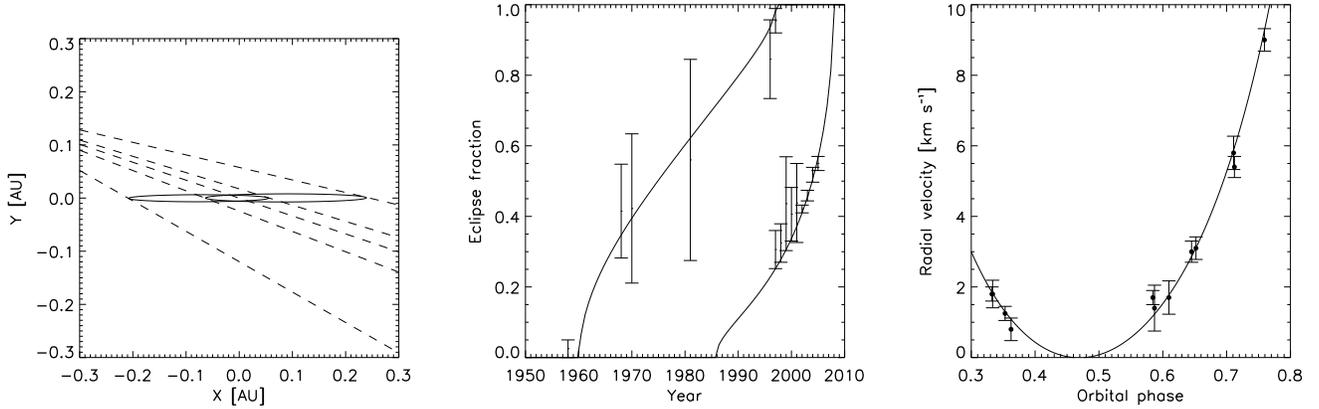}}
\end{center}
\caption{\label{fig:bestfit2}
{\bf Model 2}: $M_B/M_A = 1.15$ and $\dot{\theta}_E$ is a
free parameter. The plotting conventions are the same as in
Fig.~\ref{fig:bestfit1}.
\vspace{0.2in}
}
\end{figure*}

As explained in \S~\ref{subsec:screen}, one might expect the real
projection of a circumbinary disk to exhibit deviations from a
constant velocity. Once the screen is allowed to rotate or accelerate
during its passage in front of the binary orbit, a much wider range of
mass ratios is allowed.  While this saves the model from a seemingly
unphysical prediction, it becomes impossible to determine the stellar
mass ratio independently. Instead, in what follows, we hold the mass
ratio fixed at $M_B/M_A = 1.2$. Because the leading-order correction
in the precession angular velocity $\beta$ is a rotation, we
concentrate on the case of uniform motion plus a uniform rotation
(with no acceleration). After dropping one of the original free
parameters affecting the radial velocities ($M_B$), and adding a new
one ($\dot{\theta}_E$) affecting only the eclipses, the resulting fit
is also good, with $\chi^2 = 6$ and 5 degrees of freedom. The fit to
the data is shown in Fig.~\ref{fig:bestfit2}, and the best-fitting
parameter values are in Table~\ref{tbl:params} under the heading
``Model 2.''

We conclude that it is possible to fit the data with a physically
plausible mass ratio. All of the parameters agree with the
expectations given in \S~\ref{subsec:expectations}, including the
heliocentric radial velocity ($\gamma = 18.4$~km~s$^{-1}$).  The
best-fitting value for the rotation rate is $\dot{\theta}_E = 6\times
10^{-3}$~rad~yr$^{-1}$, which is compatible with reasonable guesses
for the circumbinary disk properties.  It implies a precession rate of
order $\beta \sim 10^{-2}$~rad~yr$^{-1}$, as would be experienced by a
circular ring of radius $\sim$3~AU.  However, a perfectly circular
ring is not realistic on physical grounds, and is not consistent with
the data beyond this order-of-magnitude comparison; the rotation rate
in that case would be exactly zero at $t_3$, in contradiction to the
assumptions of the model. We also found that a uniform acceleration
with no rotation provides a reasonable fit, and obviously a
combination of rotation and acceleration is possible. Given the
uncertainty in both the stellar mass ratio and the geometry of the
disk, we cannot use the model alone to decide which of these cases is
closer to the truth.

\section{Models of the occultations}
\label{sec:occultations}

The models described in the previous section ignore some potentially
interesting and information-bearing aspects of the photometry, such as
the phase of each individual eclipse, the shape of the ingress and
egress light curves, and the progressive deepening of the eclipses.
In this section, we elaborate upon our model in several ways and
attempt to fit all of the photometry.

In Paper~I, we generated model light curves by treating the occulting
feature as a completely opaque and semi-infinite knife-edge, and the
stars as limb-darkened photospheres.  Although these model light
curves successfully reproduce the basic patterns in the photometry,
there are at least three ways in which they fail to match the data
within the measurement uncertainties:
\begin{enumerate}
\item The model does not reproduce the observed $\sim$0.1~mag
variations outside of eclipses. As noted previously, Hamilton et al.\
(2005) provided evidence for quasiperiodic variability outside of
eclipses, and interpreted the 9.6~day period as the stellar rotation
period of star A.  Presumably, these episodic variations are caused by
cool star spots that last only a few months, as is common for young
low-mass stars (see, e.g., Herbst et al.\ 1994).
\item The steepest (central) phase of each occultation light curve is
consistent with a knife-edge crossing a photosphere.  However, during
the first and last third of each event, the flux variation is slower
than such a model would predict, as noted by Agol et al.\ (2004) and
in Paper~I.
\item Light is observed even during mid-eclipse, when (according to
the model) the photospheres of both stars are completely hidden. The
mid-eclipse flux has fallen from about 9\% to 2\% of the uneclipsed
flux over the last decade (see Fig.~\ref{fig:plotall}). Indeed, the
mid-eclipse light is not constant even within a single
eclipse. Although the re-brightening events have diminished
dramatically in intensity since 1998, there are still faint echoes of
this phenomenon in the more recent data.
\end{enumerate}

\begin{figure*}[t]
\begin{center}
  \leavevmode
\hbox{%
  \epsfxsize=6in
  \epsffile{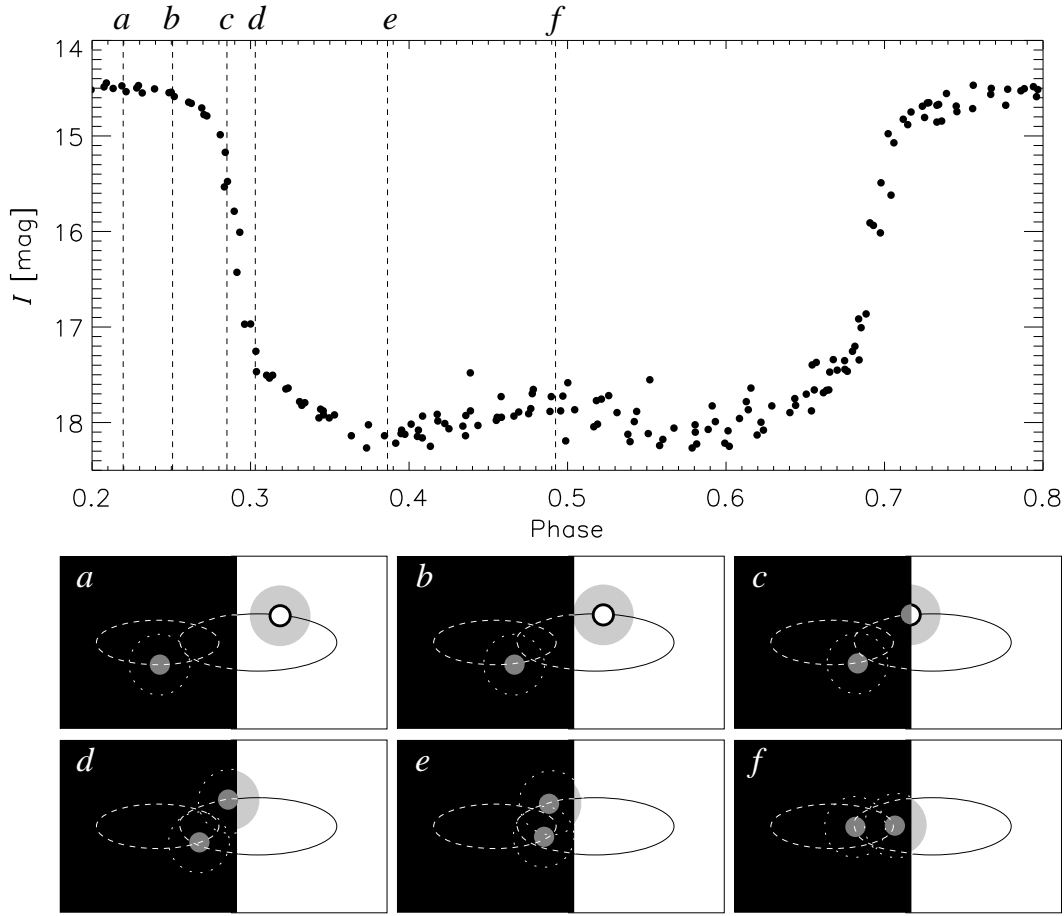}}
\end{center}
\caption{\label{fig:kink}
Illustration of the model for the occultation light curve.\
{\it Top.}---Phased light curve from the 2001--2002 season, with 6
particular phases marked with vertical dashed lines.  {\it Bottom
panels.}---Corresponding configurations of the stars, halos, and the
occulting edge. These are cartoons only, and do not represent
optimized model parameters. In particular, the best-fitting model
halos are {\it asymmetric} unlike the circular halos drawn here.
\vspace{0.2in}
}
\end{figure*}

We deal with the first point by treating the episodic variations of
star A as an intrinsic source of noise. We do not expect the light
curve to be predictable with better than 0.1~mag accuracy. As for the
second and third points, we have found that a good fit is achieved
with a simple model in which each star is enveloped by a faint halo.
Fig.~\ref{fig:kink} illustrates how this model accounts for the the
light observed during eclipses, the intra-eclipse variations, and the
shape of the ingress and egress light curves. The dashed lines mark 6
special photometric phases, labelled $a$--$f$, for which the
configurations of the photospheres, the halos, and the occulting
screen are illustrated in the smaller panels beneath the light
curve. These phases are: (a) Out of eclipse. Both the photosphere and
the halo of star A are seen. (b) Start of ingress. The halo of star A
begins to be occulted. (c) Midpoint of ingress. The photosphere is
half-covered and the total flux declines steeply. (d) The photosphere
is completely occulted, and the total flux resumes its shallower
decline. (e) The halo of star A is almost completely covered, and the
total flux reaches a local minimum.  (f) Star B reaches its closest
approach to the occulting edge, and its halo is maximally exposed. The
total flux reaches a local maximum.  (At this phase, before $t_4$, the
photosphere of star B was exposed and manifested as the re-brightening
event.) Although it is not illustrated in Fig.~\ref{fig:kink}, the
gradual darkening of the mid-eclipse phase can also be accounted for
with this model: as the occulting edge advances, the fraction of the
halo of star B that is exposed at mid-eclipse [panel (f)] gradually
shrinks.

\begin{figure}[!h]
\epsscale{1.0}
\plotone{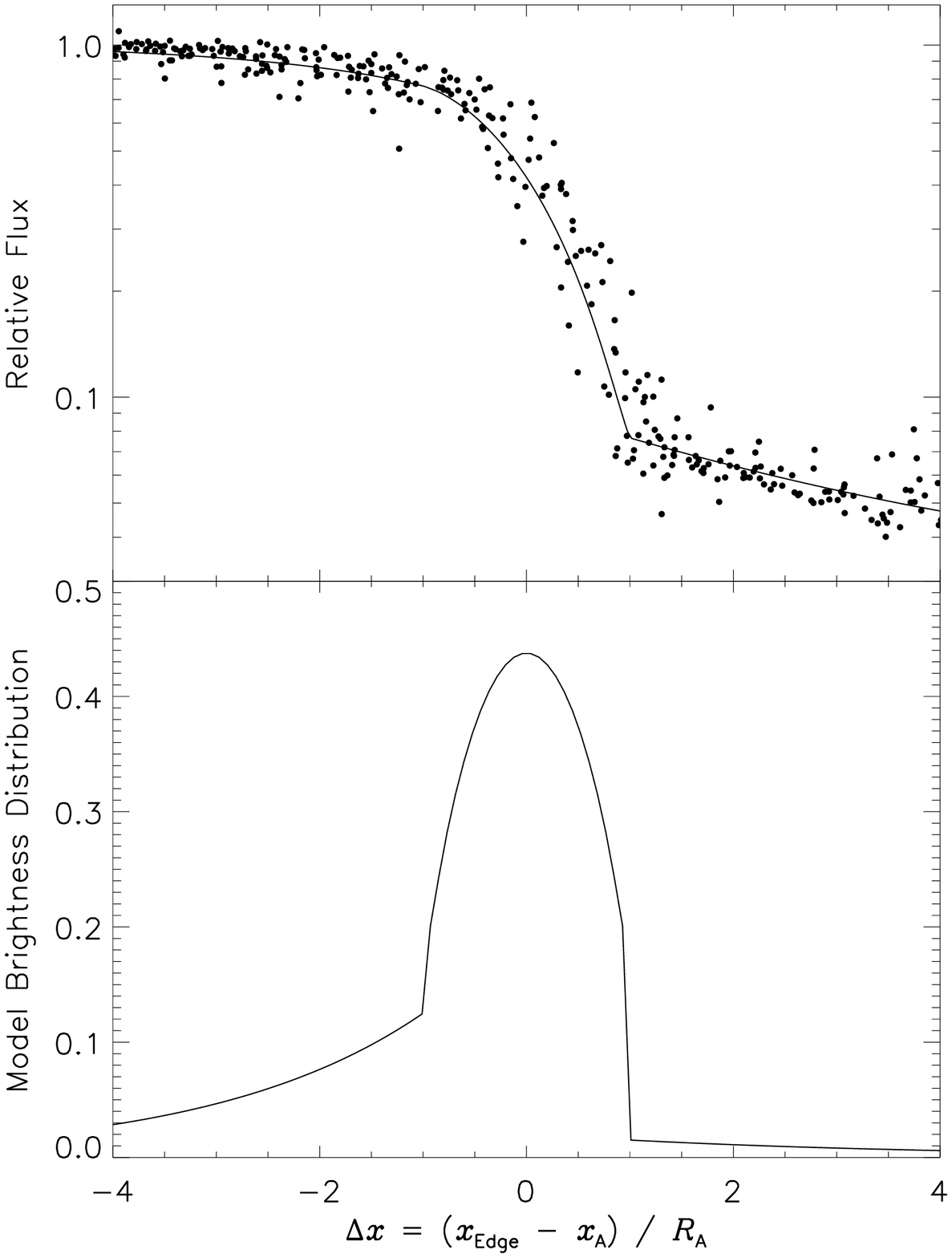}
\caption{ {\it Top}.---The relative flux of KH~15D as a function of
the position of the occulting edge.  The quantity $\Delta x$ is the
sky-projected distance between the edge and the center of star A, in
units of the stellar radius. For a perfectly sharp and straight edge,
this plot can be interpreted as the cumulative 1-d
brightness distribution of the star and its apparent halo.  The dots
show the data, and the solid line is the best-fitting model.  {\it
Bottom}.---The 1-d brightness distribution of the
best-fitting model.
\label{fig:coronagraph}}
\end{figure}

To develop a quantitative model, we first demonstrate that the
occultations of star A are {\it repeatable} events: all of the
observed ingress and egress events are nearly equivalent, after
correcting for the differences in orbital velocity of the star at the
contact points with the occulting edge. Using our model, we calculate
the position of star A and of the occulting edge as a function of
time. Then, instead of plotting the flux {\it vs}.\ time, we plot the
flux {\it vs}.\ the position of the occulting edge relative to star A.
For this purpose it is convenient to use a rotated coordinate system
$(x,y)$, in which the occulting edge is the line $x=x_E(t)$, the
perpendicular direction is the $y$-axis, and distances are measured in
units of $R_A$. The sky position of star A is $(x_A,y_A)$. The top
panel of Fig.~\ref{fig:coronagraph} shows the result, summarizing 10
years of photometry as star A goes back and forth beneath the
occulting edge. The measured flux is indeed a coherent function of
$\Delta x \equiv x_E-x_A$, with a scatter of $\sim$10\%.

This allows us to model the occultations as a knife edge scanning over
a time-invariant surface brightness distribution associated with the
star, $S(x-x_A,y-y_A)$. Since the relative velocity of the edge and
the star is always closely aligned with the $x$-axis, the only
relevant quantity is the one-dimensional brightness distribution
\begin{equation}
B(x) = \int^{\infty}_{-\infty} dy \hskip 0.05in S(x,y).
\end{equation}
When the occulting edge is at $x_E$ and star A is at $(x_A,y_A)$, the
measured flux is the cumulative 1-d brightness distribution
\begin{equation}
C = \int^{+\infty}_{x_E} dx \hskip 0.05in B(x-x_A)
  = \int^{+\infty}_{\Delta x}  dx \hskip 0.05in B(x),
\label{eq:occult}
\end{equation}
which is a function of $\Delta x$, as observed.  We note that $B$
represents all of the flux that moves with star A on the sky. It
includes light from the star, any luminous material that orbits with
the star, and also any scattering halo (which we will later argue does
exist). The halo need not be physically associated with the star, just
as the lunar halo that is sometimes observed on cold nights is not
physically associated with the Moon.

Next, we devise a fitting function for $B$ that is consistent with the
top panel of Fig.~\ref{fig:coronagraph}. At $\Delta x = 0$, the
occulting edge intersects the center of star A, and approximately
one-half of the total flux is exposed. As the screen advances over the
star from $\Delta x = 0$ to $1$, the received flux falls rapidly, in a
manner consistent with a knife edge covering a stellar
photosphere. The abrupt slope change at $\Delta x = 1$ is suggestive
of the contact between a sharp occulting edge and the rim of the
stellar photosphere. For $\Delta x > 1$, the flux falls off much more
gradually, as if the star is trailed by a spatially extended halo
comprising a few per cent of the total flux. Interestingly, the light
curve is not symmetric about $\Delta x = 0$. Once the star is more
than half exposed ($\Delta x < 0$), the flux rises more gradually than
one would expect for a symmetric halo. The side of the star nearest to
the occulting edge appears more smeared out than the other side.

After some experimentation, we found good results for the fitting
function
\begin{equation}
B(x) =
\cases{
B_1 \exp\left[ (x + 1) / \xi_1 \right], & $ x \leq -1$  \cr
B_1 + B_\star(x),                       & $-1 < x < 1$  \cr
B_2 \exp\left[-(x - 1) / \xi_2 \right], & $ x \geq  1$. \cr
}
\label{eq:brightness}
\end{equation}
The ingredients of this function are the brightness distribution of
the stellar photosphere ($B_\star$), and an exponential fall-off on
each side of the star. The amplitude ($B_1$ or $B_2$) and scale length
($\xi_1$ or $\xi_2$) of the exponential function are allowed to be
different on each side, to match the observed asymmetry in $C$, and
the function is discontinuous at $x = 1$, to match the observed slope
discontinuity in $C$. In practice, rather than quoting $B_1$ and
$B_2$, it is more useful to describe the brightness distribution by
the total flux ($L$) and the fraction of the total flux carried by
each side of the exponential halo ($\epsilon_1$ and
$\epsilon_2$). Although the photosphere may be modeled as uniform to
within the accuracy of the data, for completeness we use a linear
limb-darkening law, with a coefficient of 0.65.\footnote{For a star
  with $T_{\rm eff} \approx 4000$~K and $g = 10^4$~cm~s$^{-2}$, a
  linear limb-darkening coefficient of $u\approx 0.65$ is predicted by
  Claret (1998) and Claret (2000) for the Cousins $I$ band. Van Hamme
  (1993) predicts a similar value, 0.6. The value of 0.3 used by
  Herbst et al.\ (2002) and in Paper~I was chosen mistakenly. In any
  case, the choice of the limb-darkening coefficient has a very minor
  effect on the model light curve.} We emphasize that
Eq.~\ref{eq:brightness} is merely a fitting formula. In fact, we will
argue later that the halo is not a physical object, but rather
represents forward- or back-scattering by particles in the
circumbinary disk. In this section, we limit ourselves to
demonstrating the success of the phenomenological ``halo model,''
deferring further discussion of the physical interpretation until
\S~\ref{sec:discussion}.

Next, we proceed to the optimization of parameters. As before, we
assume $M_A = 0.6$~$M_\odot$, $R_A = 1.3$~$R_\odot$, and $M_B/M_A =
1.2$. The occultor is an opaque, semi-infinite straight edge whose
intersection point $Y_E(t)$ and rotation angle $\theta_E(t)$ are both
linear functions of time. Both star A and star B are modeled with the
function given in Eq.~(\ref{eq:brightness}), and we add a
time-invariant term $L_0$, representing any large-scale (unocculted)
components of the surface-brightness distribution. All together, the
model has $N_p=17$ free parameters: 7 specifying the
surface-brightness distribution, $\{L_A, L_B, L_0, \epsilon_1,
\epsilon_2, \xi_1, \xi_2\}$; 6 specifying the orbit, $\{P, e, I,
\omega, T_p, \gamma\}$, and 4 describing the occulting screen, $\{t_4,
t_5, \theta_E(t_4), \dot{\theta}_E\}$.

It was desirable to speed up the computations by time-averaging the
photometry.  Whenever more than one measurement was made within a
given 6~hour time period, we computed the flux-weighted mean of the
results. The resulting binned data set has $N_f = 1044$ entries. The
fitting statistic is
\begin{equation}
\chi^2 =
  \sum_{i=1}^{N_f} \left( \frac{ f_{C,i} - f_{O,i} } {\sigma_{f,i}} \right)^2 +
  \lambda \sum_{i=1}^{N_V} \left( \frac{ V_{C,i} - V_{O,i} } {\sigma_{V,i}} \right)^2,
\label{eq:chi2}
\end{equation}
which is analogous to Eq.~(\ref{eq:chi2rv}), with the fluxes $f$
replacing the ingress or egress durations $d$. The uncertainty
$\sigma_{f,i}$ was taken to be the quadrature sum of the measurement
uncertainty and 10\%, where the latter is intended to account for
intrinsic variations. The multiplier $\lambda$ controls the relative
weighting of the flux data and radial velocity data. If the
measurement uncertainties were known perfectly, and the model were
correct, then $\lambda=1$ would be the appropriate choice and the
best-fitting solution would have $\chi^2 \approx 1040 = N_f + N_V -
N_p$.

In this case, $\lambda=1$ is probably not the best choice. We expect
the model to be statistically consistent with the radial velocity
data, because the only relevant aspect of the model---the assumption
of a fixed Keplerian orbit---seems reasonable. In contrast, the
photometric model involves idealized assumptions about the occultor
and the surface-brightness distribution, and the estimates of
$\sigma_f$ are crude. Our approach was to force the model to be
statistically consistent with the radial velocity data, by increasing
$\lambda$ until the best-fitting solution had $\chi^2_V \approx 12 =
N_V$. Of course, we cannot be certain that the radial-velocity data
are free from additional sources of measurement error and unmodeled
systematic effects, such as precession of the orbit due to
perturbations from the circumbinary disk and additional
bodies. Nevertheless, in the rest of this section, we describe the
optimized model that results from this procedure.

\begin{figure*}[t]
\begin{center}
  \leavevmode
\hbox{%
  \epsfxsize=7in
  \epsffile{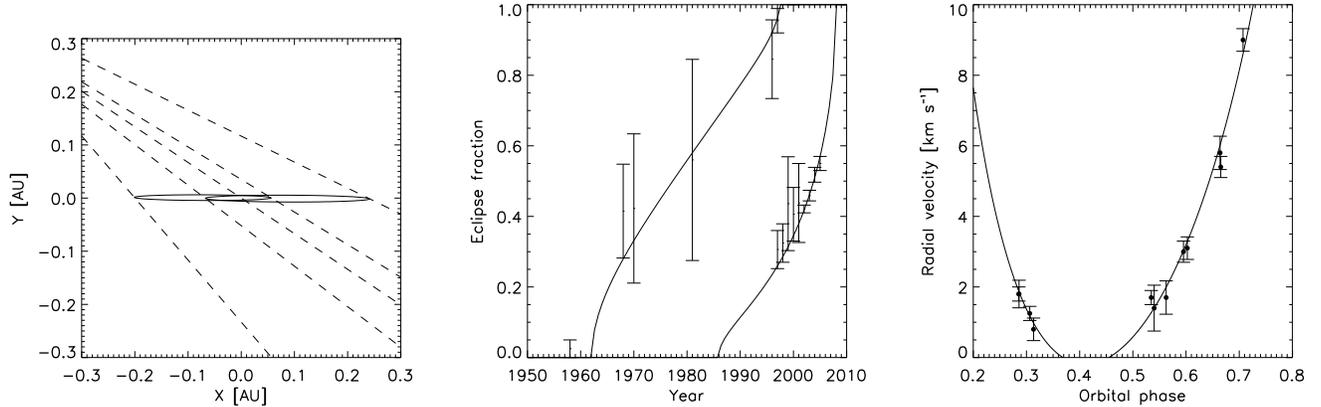}}
\end{center}
\caption{\label{fig:bestfit3}
{\bf Model 3}: $M_B/M_A = 1.2$, $\dot{\theta}_E$ is a free
parameter, and the full photometric time series is fitted (rather than
just the eclipse fractions and partial-eclipse durations).  The
plotting conventions are the same as in Fig.~\ref{fig:bestfit1}.
\vspace{0.2in}
}
\end{figure*}

\begin{figure*}[b]
\begin{center}
  \leavevmode
\hbox{%
  \epsfxsize=7in
  \epsffile{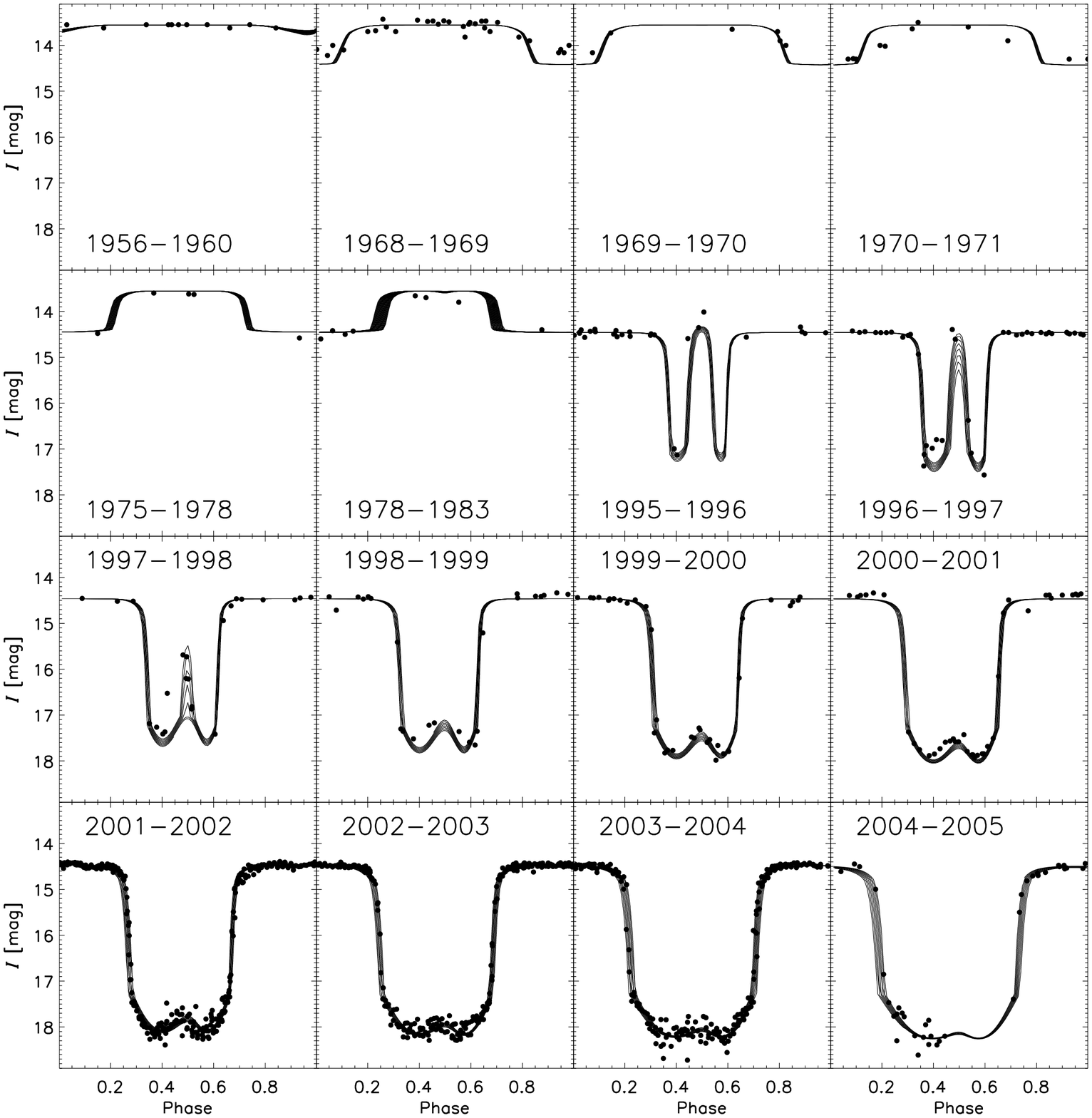}}
\end{center}
\caption{\label{fig:bestfit3_lc}
{\bf Model 3}: Calculated and observed light curves. Each panel
shows the phased light curve from a particular time range, as observed
(solid symbols) and calculated (dots).  The boundaries of the time
ranges are marked with dashed lines in Fig.~\ref{fig:plotall}.
\vspace{0.2in}
}
\end{figure*}

With $\lambda=50$, the best-fitting model has $\chi^2_V = 13$ and
$\chi^2_f = 1574$. The orbit, and the fit to the radial velocity data,
are shown in Fig.~\ref{fig:bestfit3}. The fit to the photometry is
shown in Fig.~\ref{fig:bestfit3_lc}. Each panel shows the phased light
curve for a particular time interval. The time intervals chosen for
this plot are the intervals that are marked by dashed lines in
Fig.~\ref{fig:plotall}. The final column of Table~\ref{tbl:params},
under the heading ``Model 3,'' gives the best-fitting parameters.

\begin{deluxetable}{cccc}
\tabletypesize{\small}
\tablecaption{KH~15D Model Parameters}
\tablewidth{0pt}
\tablehead{
\colhead{Parameter} &
\multicolumn{3}{c}{Best-fitting Value} \\
\colhead{} &
\colhead{Model 1} &
\colhead{Model 2} &
\colhead{Model 3}
}
\startdata
                  $M_A$ [$M_\odot$]  & $   0.6   $ & $   0.6        $ & $   0.6\pm 0.1         $ \\
                          $M_B/M_A$  & $   0.76  $ & $   1.2        $ & $   1.2\pm 0.1         $ \\
                  $L_A$ [$L_\odot$]  &    \nodata   &     \nodata     & $   0.413\pm 0.006     $ \\
                          $L_B/L_A$  &    \nodata   &     \nodata     & $   1.36\pm 0.10       $ \\
                          $L_0/L_A$  &    \nodata   &     \nodata     & $   0.029\pm 0.005     $ \\
                  $R_A$ [$R_\odot$]  & $    1.3   $ & $   1.3       $ & $   1.30\pm 0.07       $ \\
                          $R_B/R_A$  & $    1.05  $ & $   1.05      $ & $   1.05\pm 0.03       $ \\
                            $\xi_1$  &    \nodata   &     \nodata     & $   2.0\pm 0.3         $ \\
                            $\xi_2$  &    \nodata   &     \nodata     & $   3.3\pm 0.4         $ \\
                       $\epsilon_1$  &    \nodata   &     \nodata     & $   0.25\pm 0.05       $ \\
                       $\epsilon_2$  &    \nodata   &     \nodata     & $   0.049\pm 0.003     $ \\
                         $P$ [days]  & $   48.4   $ & $   48.4      $ & $   48.381\pm 0.005    $ \\
                                $e$  & $    0.55  $ & $    0.58     $ & $   0.574\pm 0.017     $ \\
                          $I$ [deg]  & $   83     $ & $   83        $ & $  92.5 \pm 2.5        $ \\
                     $\omega$ [deg]  & $    6     $ & $   14        $ & $   13\pm 2            $ \\
         $T_p$~[JD]~$-$~2,452,350    & $    1.1   $ & $    0.7      $ & $  1.9 \pm 0.7         $ \\
             $\gamma$ [km~s$^{-1}$]  & $   14.1   $ & $   18.4      $ & $  18.6 \pm 1.3        $ \\
                              $t_1$  & $ 1965     $ & $ 1963        $ & $ 1962 \pm 3           $ \\
                              $t_2$  & $ 1985.3   $ & $ 1985.9      $ & $ 1985.9 \pm 0.7       $ \\
                              $t_3$  & $ 1991.1   $ & $ 1992.5      $ & $ 1992.7 \pm 0.5       $ \\
                              $t_4$  & $ 1998.8   $ & $ 1997.0      $ & $ 1997.3 \pm 0.5       $ \\
                              $t_5$  & $ 2010.9   $ & $ 2008.0      $ & $ 2008.0 \pm 0.2       $ \\
              $\theta_E(t_4)$ [deg]  & $  -20     $ & $  -16        $ & $ -32 \pm 7            $ \\
   $\dot{\theta}_E$ [rad~yr$^{-1}$]  & $    0     $ & $ 0.0061      $ & $ 0.0087 \pm 0.0015    $ 
\enddata

\tablecomments{Models 1 and 2 were fitted to the eclipse duration,
  ingress/egress duration, and radial velocity data.  In Model 1,
  $M_B$ is a free parameter and $\dot{\theta}_E = 0$.  In Model 2,
  $M_B/M_A = 1.2$ and $\dot{\theta}_E$ is a free parameter.  Model 3
  was fitted to the photometric and radial velocity data. The
  uncertainties were determined with the Monte Carlo procedure
  described in the text.}

\label{tbl:params}
\end{deluxetable}

The table includes estimated uncertainties in the fitted quantities,
which should be treated with caution. They were calculated using a
Monte Carlo algorithm, as follows. We optimized the parameters on each
of $10^4$ synthetic data sets (``realizations''). Each realization
consisted of $N_f=1044$ flux measurements and $N_V=12$ radial velocity
measurements (the same numbers as the real data set). The flux
measurements were randomly selected from the real data set, with
repetitions allowed. The intent is to estimate the probability
distribution of the measurements using the measured data values
themselves (see, e.g., Press et al.\ 1992, p.\ 689). The number of
radial velocity measurements is too small for drawing-with-replacement
to be effective. Instead, we added Gaussian random numbers to the
measured radial velocities, with a mean of zero and a standard
deviation given by the quoted $\sigma_{v}$ values. To account for the
uncertainty in the stellar masses, we assigned $M_A/M_\odot$ by
picking a random number from a Gaussian distribution with mean $0.6$
and standard deviation $0.1$. Likewise, for the mass ratio $M_B/M_A$,
we used a Gaussian distribution with a mean of $1.2$ and a standard
deviation of $0.1$. The stellar radii were fixed according to the
relation $R\propto M^{0.27}$. For each parameter, we found the median
and the approximate 68\% confidence limits in the resulting
distribution of best-fitting values.

There are many caveats. To begin with, the model is not in formal
statistical agreement with the data (some discrepancies are
highlighted below). The random flux variations of the stars are not
independent; they are correlated on time scales of days. The {\it a
  priori} probability distributions for the stellar masses are
probably not Gaussian. The systematic error related to our assumption
about the trajectory of the occulting edge has not been taken into
account. One can imagine more complex analyses that attempt to
overcome these problems, but we do not believe that such methods are
justified.

The model shares the successes of the models presented in
\S~\ref{sec:orbit}. It reproduces the transitions from no eclipses to
partial eclipses to total eclipses, the existence of the central
reversals from at least 1995 until 1999, and the growing duty cycle of
the modern eclipses. The more detailed photometric model also accounts
for the relative phases of the partial and total eclipses, the gradual
deepening of the eclipses and the slope discontinuities in the ingress
and egress light curves. In addition to fitting the data reasonably
well, the model satisfies all of the criteria that were discussed in
\S~\ref{subsec:expectations}. The stellar masses, radii, and
luminosities are reasonable, by fiat. The orbital eccentricity is
smaller than the theoretical upper bound of 0.66, and the orbital
inclination is close to edge-on. The heliocentric radial velocity of
the center of mass is near the median that has been observed among
cluster members. The screen trajectory is consistent with
order-of-magnitude estimates for the expected motion of the edge of a
precessing circumbinary ring with a radius of $\sim$3~AU.

The best-fitting brightness distribution for star A is shown in
Fig.~\ref{fig:coronagraph}. The upper panel shows the model cumulative
1-d brightness distribution overlaid on the data points, while the
lower panel shows the model brightness distribution itself. The
asymmetry of the halo is evident. The exponential scale length on both
sides is a few stellar radii, but the fractional flux of the halo is
much greater on the side closest to the occulting edge ($\sim$25\%)
than on the other side ($\sim$5\%).

As noted before, the model cannot reproduce the apparently random
$\sim$10\% flux variations of star A that occur on time scales of a
few days.  Nor can it reproduce the seasonal fluctuations of a few per
cent in the mean flux of star A. There are some other significant
discrepancies between the data and the model. In the 1967-1972 time
period, there is a $\sim$0.1~mag offset between the calculated and
observed light curves, and in addition, the calculated eclipse depth
is $\sim$0.15~mag too deep. These problems are at least partly the
fault of the uncertainty in the photographic magnitude scale. The
offset is within the 0.14~mag uncertainty in the zero point, and the
calculated depth becomes too {\it shallow} if we use the $I_M$ system
instead of the $I_J$ system (see \S~\ref{subsec:plates}). Another
discrepancy is the phase shift of $\sim$0.02 between some of the
calculated and observed central re-brightenings. This is most easily
seen in the 2000--2001 data. This offset disappears for an orbital
period of 48.36~days, but that value of $P$ is disfavored by the
radial velocity data and by the relative phases of the pre-1980
eclipses. Finally, there is one data point in 1997--1998 that is very
poorly described: at $I=16.5$, it would seem that the photosphere of
star B was partly exposed, but in the model, it was behind the screen.

\vspace{0.5in}
\section{Summary and Discussion}
\label{sec:discussion}

Based on the preceding results, we draw two conclusions. First, the
newly-available data corroborates the theory that KH~15D is an
eccentric binary system being occulted by the edge of a precessing
circumbinary disk. As shown in \S~\ref{sec:orbit}, the model presented
in Paper~I succeeds as a quantitative description of KH~15D, even when
confronted with far more data than were available when the model was
invented. The best-fitting parameter values are realistic and conform
to theoretical expectations. A problem that was overlooked in
Paper~I---the seemingly unphysical mass ratio of the stars---is fixed
with a minor and physically-motivated elaboration to the model,
namely, the allowance for a more realistic trajectory for the
occulting edge across the line of sight. Second, the occultation light
curves are well-fitted by a model in which each star is surrounded by
a more extended halo. This model succinctly accounts for both the
photometric variations observed during individual events, and the
gradual deepening of the eclipses. The halo around each star is not
symmetric about the center of the star, but in each direction it has a
typical scale length of a few stellar radii, and it is brighter in the
direction facing the occulting edge. Only the one-dimensional profile
of the halo has been measured.

What are these halos? The three broad categories of possible physical
interpretations of each halo are (1) {\it luminous} material, such as
accretion columns or a hot corona, (2) partially {\it extinguished}
starlight, (3) {\it scattered} starlight. The evidence does not permit
an unambiguous interpretation, but it does strongly suggest that the
halo light has a scattered component. Relative to the uneclipsed
light, the mid-eclipse light has a larger fractional polarization
(Agol et al.\ 2004) and is slightly bluer in color (Hamilton et al.\
2001, 2005; Herbst et al.\ 2002; Knacke, Fajarado-Acosta \& Tokunaga
2004; Kusakabe et al.\ 2005).  Enhanced polarization and blueness are
hallmarks of scattering by small particles. In contrast, partially
extinguished light would be {\it redder} than the light source (or the
same color, if the particles causing the extinction were large
enough). Likewise, the observation of only a small color change argues
against self-luminous material in an accretion flow or a hot corona,
which would not be expected to have nearly the same effective
temperature as the photosphere.

\begin{figure}[!h]
\epsscale{1.0}
\plotone{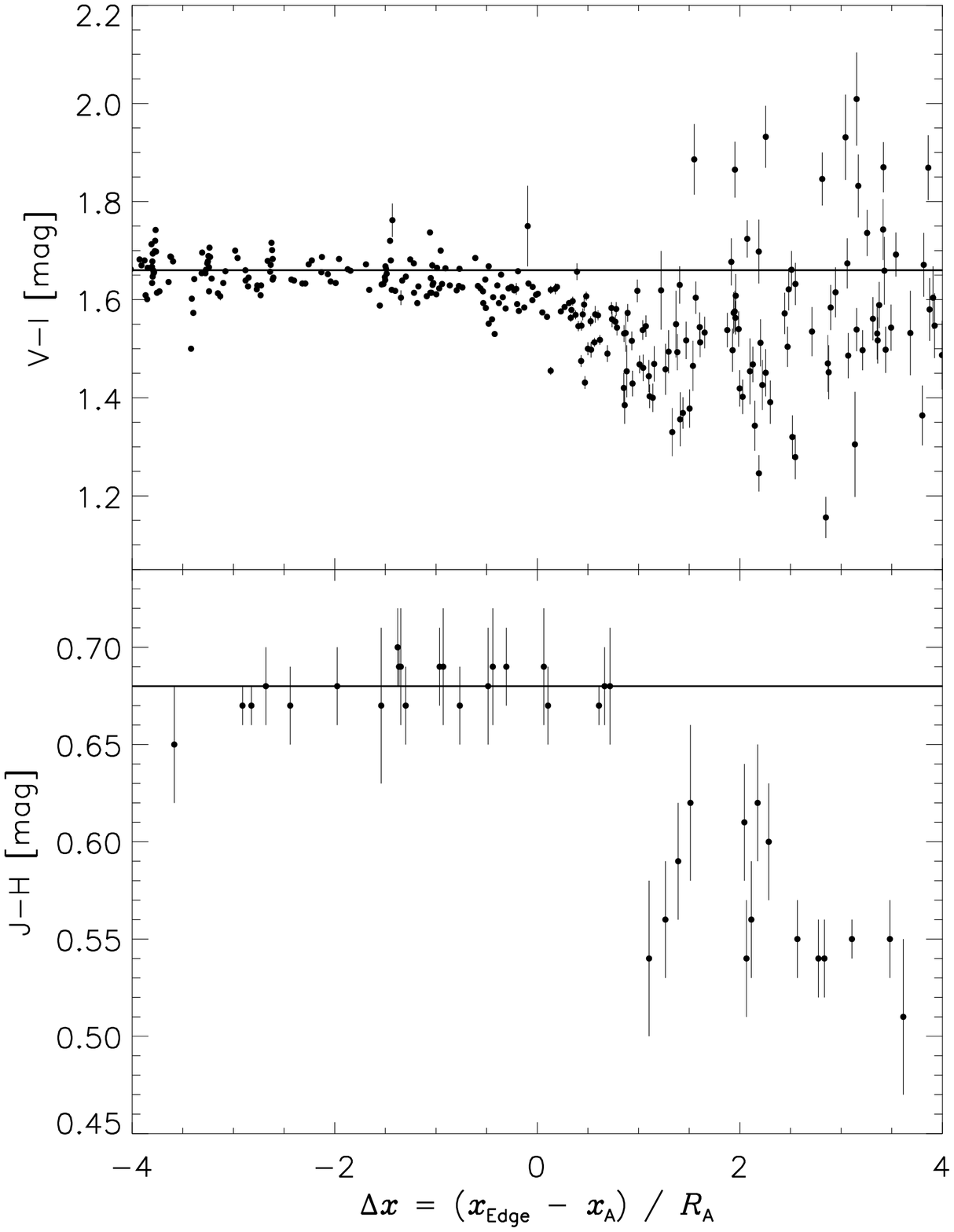}
\caption{ The color index of KH~15D, as a function of the distance
between the occulting edge and the center of star A. The median color
out-of-eclipse is indicated with a solid line. To make this plot, the
time stamps on the $V-I_J$ photometry of Hamilton et al.\ (2005) and
the $J-H$ photometry of Kusakabe et al.\ (2005) were converted into
$(x_E - x_A)$, using the model presented in
\S~\ref{sec:occultations}. The halo is seen to be bluer than the
photosphere at both optical and near-infrared wavelengths.
\label{fig:coronagraph_color}}
\end{figure}

Just as the {\it brightness} distribution of the halo was determined
from $I$-band photometry in \S~\ref{sec:occultations}, the {\it color}
distribution of the halo can be measured via multi-band photometric
monitoring. A first attempt at this is shown in
Fig.~\ref{fig:coronagraph_color}, which is analagous to
Fig.~\ref{fig:coronagraph} but uses the multi-band photometry of
Hamilton et al.\ (2005) and Kusakabe et al.\ (2005) rather than
monochromatic photometry. Higher-precision monitoring in multiple
bands throughout an ingress or egress will help to resolve the color
structure of the halo.

As for the location of the scattering material, the two basic
possibilities are the immediate vicinity of each star, and the
circumbinary disk. Examples of the first scenario are scattering from
a corona of hot electrons, or from infalling material. A difficulty
with this scenario is the near-repeatability of the occultation light
curve. A circumstellar halo would need to maintain a fixed orientation
in space with respect to the occulting feature, despite the rotation
of the star and its tidal interaction with its binary companion. Disk
scattering, on the other hand, would need to be dominated by
forward-scattering or back-scattering (or some combination of the
two), in order to give the appearance of a localized halo around the
star. The more general illumination of the disk would not depend
strongly on the positions of the stars, because the disk radius of
$\sim$3~AU is much larger than the binary semimajor axis of
$\sim$0.25~AU. It is possible, however, that this is a false
dichotomy. The disk may extend inward continuously to much smaller
distances, and indeed the strong H$\alpha$ emission, indicative of
ongoing accretion, can be taken as evidence for this.

One way in which these possibilities might be distinguished is through
the difference in the typical scattering angle. In the case of
circumstellar scattering, the received light is scattered over a wide
range of angles, as opposed to the forward- or back-scattering from
the disk. The chromatic and polarization signals associated with
scattering depend on the typical scattering angle, as well as the size
and composition of the particles. Hence, measurements of the polarized
light curve, combined with models of the expected polarization signal
under various circumstances, may help to determine the size,
composition and spatial distribution of the scattering particles. It
would also be possible to determine the velocity of the scattering
material relative to the stars, through observations of Doppler shifts
in the stellar absorption features of the scattered light.

Another interesting feature of the model is the extreme sharpness of
the occulting edge. The best evidence for a sharp edge is the slope
discontinuity in the light curve that was pointed out in
\S~\ref{sec:occultations}. This ``kink'' is most easily understood as
the point where the sharp occulting edge contacts the rim of the
photosphere of the star. If the edge were not sharp, then the time of
photospheric contact would be smeared out. The presence of the kink
requires that the optical depth of the edge increases from nearly zero
to $>$3.5 over a distance that is smaller than a stellar radius. It
can also be argued that the edge is sharp on the scale of a stellar
radius based on the dramatic changes in the H$\alpha$ emission and
absorption profile that Hamilton et al.\ (2003) observed during an
occultation. The implication is that the occulting edge spatially
resolves the H$\alpha$-emitting region, which is thought to arise
within a few stellar radii (i.e.\ within the magnetosphere).

An independent way to assess the sharpness of the edge would be to
measure the Rossiter-McLaughlin effect. In \S~\ref{sec:rv}, we
predicted the amplitude of this effect for a perfectly sharp and
straight edge. In the other extreme, in which the scale length of the
edge is much larger than the stellar radius, there is no Rossiter
effect. Thus, a sequence of optical spectra with a high
signal-to-noise ratio throughout an occultation could be used to
determine whether or not the occulting edge resolves the stellar
photosphere. If the occulting edge is truly sharp on the scale of the
stellar radius, then it can be used to map out the environment of star
A on scales unobservable by any other means, just as lunar
occultations were used to identify radio and X-ray sources before the
necessary angular resolution was available. Spectroscopic monitoring
would resolve the velocity structure of the surrounding material. The
projected orbital velocity of star A relative to the edge is
approximately one stellar radius per day. Thus, a time sampling of 1
hour corresponds to a spatial resolution of $\sim$0.05 stellar radii,
or an angular resolution of $\sim$0.5~$\mu$as at a distance of
760~pc. However, the scattering geometry will need to be understood
first, to allow the information in the halos to be decoded.

An interesting theoretical question is {\it why} the edge is sharp.
With a radius of $a_r = 3$~AU and a vertical scale height $h <
R_\odot$, the circumbinary disk would have $h/a_r < 10^{-3}$. As
pointed out by Chiang \& Murray-Clay (2004), this is at least an order
of magnitude smaller than the values generally assumed for T~Tauri
stars, which are based on the consideration of hydrostatic equilibrium
between gas pressure in the disk and the vertical gravitational field
provided by the star. One possible resolution is that there is little
or no gas over the range of orbital distances that form the occulting
feature. Another possibility is that $a_r$ is smaller than the
dynamical arguments suggest. A third and most intriguing possibility
is that the obscuring particles in the disk have settled to the
midplane and formed a much sharper layer than the distribution of
gas. The dust-settling process has long been investigated as a
possible mechanism for generating planetesimals, by enhancing the
surface density of solid particles until a gravitational instability
ensues (Safronov 1969, Goldreich \& Ward 1973, Weidenschilling 1977).

Finally, we end this discussion on a note of urgency. A problem with
all of the observations proposed above is that the edge is continuing
to advance and will eventually cover star A at all orbital
phases. According to the model of \S~\ref{sec:occultations}, the
photosphere of star A will set for the last time near the beginning of
2008, although that prediction hinges on our extrapolation of the
trajectory of the occulting screen. The halo of star A will be visible
for several years after the photosphere is hidden. We encourage
continued monitoring and will gladly provide predictions of future
occultations to interested parties.

\acknowledgments We are grateful to V.\ Grinin and O.\ Yu.\ Barsunova
for providing their optical data, and to M.\ Tamura for providing his
group's near-infrared data. We thank J.\ Barranco, E.\ Ford, S.\
Gaudi, S.\ Kenyon, R.\ Narayan, and G.\ Rybicki for helpful
discussions. S.\ Gaudi, V.\ Grinin, and the anonymous referee provided
constructive criticism of the manuscript.

\end{document}